\documentclass[12pt]{article}

\usepackage{amsmath}
\usepackage{graphicx}
\usepackage{amsfonts}
\usepackage{amssymb}
\usepackage{epsfig}
\usepackage{color}
\usepackage{psfrag}

\newcommand{\EQ}{\begin{equation}}
\newcommand{\EN}{\end{equation}}

\begin{document}
\setcounter{page}{0} \topmargin 0pt
\renewcommand{\thefootnote}{\arabic{footnote}}
\newpage
\setcounter{page}{0}

\begin{titlepage}

\begin{flushright}
ISAS/32/2004/FM
\end{flushright}

\vspace{0.5cm}
\begin{center}
{\Large {\bf Semiclassical Scaling Functions }}\\
{\Large {\bf of Sine--Gordon Model}} \\

\vspace{2cm}
{\large G. Mussardo$^{1,2}$, V. Riva$^{1,2}$ and G. Sotkov$^{3}$} \\
\vspace{0.5cm} {\em $^{1}$International School for Advanced Studies}\\
{\em Via Beirut 1, 34100 Trieste, Italy} \\
\vspace{0.3cm} {\em $^{2}$Istituto Nazionale di Fisica Nucleare}\\
{\em Sezione di Trieste}\\\vspace{0.3cm} {\em $^3$ Instituto de Fisica Teorica} \\
{\em Universidade Estadual Paulista} \\
{\em Rua Pamplona 145, 01405-900 Sao Paulo, Brazil}

\end{center}
\vspace{1cm}

\begin{abstract}
\noindent We present an analytic study of the finite size effects
in Sine--Gordon model, based on the semiclassical quantization of
an appropriate kink background defined on a cylindrical geometry.
The quasi--periodic kink is realized as an elliptic function with
its real period related to the size of the system. The stability
equation for the small quantum fluctuations around this classical
background is of Lam\'e type and the corresponding energy
eigenvalues are selected inside the allowed bands by imposing
periodic boundary conditions. We derive analytical expressions 
for the ground state and excited states scaling functions, which
provide an explicit description of the flow between the IR and UV
regimes of the model. Finally, the semiclassical form factors and
two-point functions of the basic field and of the energy operator
are obtained, completing the semiclassical quantization of the 
Sine--Gordon model on the cylinder.

\vspace{2cm}

\hrulefill

E-mail addresses: mussardo@sissa.it,\,\, riva@sissa.it,\,\,
sotkov@ift.unesp.br

\end{abstract}

\end{titlepage}

\newpage

\section{Introduction}\label{intro}
\setcounter{equation}{0}

Quantum field theory on a finite volume is a subject of
both theoretical and practical interest. It almost invariably
enters the extrapolation procedure of numerical simulations,
limited in general to rather small samples, but it is also
intimately related to quantum field theory at finite temperature.
It is therefore important to increase our ability in treating
finite size effects by developing efficient analytic means. In the
last years, a considerable progress has been registered in
particular on the study of finite size behaviour of two
dimensional systems. Also for these models, however, an exact
treatment of their finite size effects has been obtained only in
particular situations, namely when the systems are at criticality
or if they correspond to integrable field theories. At criticality, 
in fact, methods of finite size scaling and Conformal
Field Theory \cite{finitesize,cardy} permit to determine many
universal amplitudes and to extract as well useful information 
on the entire spectrum of the transfer matrix. Away from criticality,
exact results can be obtained only for those integrable theories
described by a factorized and elastic scattering matrix
\cite{zamzam,gmrep} which, on a finite volume, can be further
analysed by means of Thermodynamical Bethe Ansatz [5--9]. 
This technique provides integral equations for the energy levels,
mostly solved numerically. In all other cases, the control of 
finite size effects in two dimensional QFT has been reached up to now 
either by conformal perturbation theory or numerical
methods as, for instance, the one proposed in \cite{TCSA}.

The aim of this paper is to study the finite size effects of a two 
dimensional massive QFT by using a different approach, i.e. the 
non--perturbative semiclassical expansion formulated in the infinite 
volume case by Dashen, Hasslacher, Neveu \cite{DHN} and by Goldstone 
and Jackiw \cite{goldstone}. Apart from some issues which make such an 
analysis an interesting subject in itself, the main theoretical motivation 
of this work consists in the possibility of obtaining analytic results for the 
form factors and the energy levels at a finite geometry. In integrable cases, 
this adds to the above techniques (see also \cite{smirnov}), whereas 
for non--integrable models it is an efficient alternative to perturbative or 
numerical studies. As a matter of fact, in the infinite volume case, semiclassical 
methods have proved to be, together with Form Factor Perturbation Theory \cite{ffpt}, ideal
tools in the analysis of non--integrable quantum field theories
(see, for instance, Ref.\,\cite{dsgmrs}). 

Form factors at a finite volume of local operators in both integrable and 
non--integrable theories have been studied in one of our previous papers 
\cite{finvolff}. These quantities 
enter the spectral density representation of correlation functions which need, 
however, another set of data for their complete determination, precisely the 
energies of the intermediate states at a finite volume. This paper is mainly 
devoted to fill this gap, that is, to face the problem of a semiclassical 
computation of the energies $E_i(R)$ of vacua and excited states as functions 
of the circumference $R$ of a cylindrical geometry. Notice that, isolating a 
factor $1/R$ in front of the $E_i(R)$'s (simply due to their dimensionality), the 
remaining quantities are {\em scaling functions} of the variable $r = m R$,
where $m$ is the lowest mass of the considered QFT.

It is worth to underline an important feature that has come out
from the study of the semiclassical form factors in infinite
volume. As we will discuss later, their accuracy seems to extend,
somehow, beyond the regime in which they were supposed to be
valid. Together with the known fast convergency properties of the
spectral series and the information that can be extracted on
energy levels, the above result suggests that the semiclassical
method may provide a rather precise estimate of finite volume
correlation functions, an outcome which may be useful for many
applications.

For methodological reasons, we have decided to present the
semiclassical computation of finite volume energies for a system 
that admits one of the simplest analysis, the Sine--Gordon (SG)
model. As we will see, this model is particularly appealing for
its simplified semiclassical results whereby the significant
physical effects we are looking for will not be masked by other
additional complications. Moreover, due to the integrable nature
of this theory, its finite size effects have been previously
studied by means of Thermodynamical Bethe Ansatz \cite{ddv,RavaniniSG}, 
and it would be interesting to perform a quantitative comparison 
between these results and the semiclassical ones, in order to 
directly control their range of validity. However, as already 
pointed out, semiclassical methods apply not only to integrable 
theories and this opens the way to describe analytically the 
finite size effects also in non--integrable models \cite{preparation}.

The paper is organized as follows: in Section \ref{sectsm} we
briefly recall the main ideas and results of the semiclassical
approach. We also discuss the simplest scaling function in a
finite volume in order to clarify the nature of divergencies
encountered in such computations. Section \ref{sectscaling} is
devoted to the complete semiclassical analysis of the energies of
the quantum states in the kink sector of the SG model on a
cylinder. In general, this analysis passes through the solution of
a Schr\"{o}dinger type equation for a particle in a periodic
potential and, for the SG model, this corresponds to a Lam\'e
differential equation. In this section we also discuss how to
select the proper eigenvalues inside the band structure of the spectrum
in order to determine the energy levels $E_i(R)$. In Section \ref{sectff}
we compute the form factors of local operators by using the
semiclassical methods and we comment on their properties. Our
conclusions and further directions are discussed in Section
\ref{conclusions}. There are also several appendices: Appendix A
presents the quantization of a free bosonic theory in a finite
volume and a comparison of finite--volume and finite--temperature
computations of the simplest one--point correlation function.
Appendix B collects relevant mathematical properties of the
elliptic functions used in the text whereas Appendix C displays
the main properties of the Lam\'e equation.

\section{Semiclassical quantization}\label{sectsm}

In this section, after recalling the basic equations of the
semiclassical quantization, we will present the simplest example
of a scaling function on a cylindrical geometry, i.e. the ground
state energy ${\cal E}_0^{\text{vac}}(R)$ of a free massive bosonic 
field. In a semiclassical quantization, ${\cal E}_0^{\text{vac}}(R)$ 
is the lowest energy level in the vacuum sector of the theory. 
This example will show, in particular, how to handle the divergencies 
usually encountered in the calculation of the scaling functions.

\subsection{DHN method}\label{mainidea}
The main feature of a large class of 2-D field theories with non--linear interaction and
discrete degenerate minima is that they admit non--perturbative finite--energy classical
solutions (called kinks or solitons) carrying topological charges $Q_{\textrm{top}}^{\pm}
= \pm 1$. In this paper we will concentrate our attention, in particular, on a specific
model of this kind, i.e. the Sine--Gordon (SG), defined by the potential
\begin{eqnarray}
V_{SG}(\phi) & = & \frac{m^2}{\beta^2} \left(1- \cos\beta \phi\right) \,\,\,.
\label{VSG}
\end{eqnarray}
In such theories, the kinks generally interpolate between two next
neighbouring minima of the potential (vacua) which are constant
solutions of the equation of motion (in our example $\phi_{SG} =
\frac{2 \pi s}{\beta}$, $s=0,1$), and they exhibit certain
particle properties. For instance, they are localized and
topological stable objects, i.e. they do not decay into mesons
with $Q_{\textrm{top}} = 0$. Moreover, in integrable theories as SG
model, their scattering is dispersionless and, in the collision
processes, they preserve their form simply passing through each
other.

The kinks are static solutions of the equation of motion, i.e. they are time
independent in their rest frame, and they can be simply obtained by integrating
the first order differential equation
\begin{equation}
\frac{1}{2}\left(\frac{\partial \phi_{cl}}{\partial x}\right)^{2}
\,=\, V(\phi_{cl})
+ A \,\,\,,
\label{firstorder}
\end{equation}
further imposing that $\phi_{cl}(x)$ reaches two different minima
of the potential $V(\phi)$ at $x \rightarrow \pm \infty$. These
boundary conditions, which describe the infinite volume case,
require the vanishing of the integration constant $A$. As we will
see in the next Section, the kink solutions in a finite volume
correspond instead to a non--zero value of $A$, related to the
size $R$ of the system.

All the above properties of the kink solutions are an indication
that they can survive the quantization, giving rise to the quantum
states in one--particle sector of the corresponding QFT. A direct
correspondence among the kink states and the corresponding
classical solution has been established by Goldstone and Jackiw
who have shown in Ref.\,\cite{goldstone} that the matrix element
of the field $\phi$ between kink states is given, at leading order
in the semiclassical limit, by the Fourier transform of the kink
background. We will discuss this result and its applications in
Sect.\,\ref{sectff}.

At the moment we are mainly concerned with the semiclassical
quantization of the small fluctuations around kink backgrounds.
As it is well known, one cannot apply directly to them the
standard perturbative methods of quantization around the free
field theory since the kinks are entirely non--perturbative solutions
of the interacting theory. Their classical mass, for instance, is
usually inversely proportional to the coupling constant (in the SG
model one has $M_{cl} = \frac{8 m}{\beta^2}$). In an infinite
volume, an effective method for the semiclassical quantization of
such kink solutions (as well as of the vacua ones) has been
developed in a series of papers by Dashen, Hasslacher and Neveu
(DHN) \cite{DHN} by using an appropriate generalization of the WKB
approximation in quantum mechanics (see also Refs. \cite{pathint}
or \cite{raj} for a review). The DHN method consists in initially
splitting the field $\phi(x,t)$ in terms of its classical static
solution of the equation of motion (which can be either one of the
vacua or the kink configuration) and its quantum fluctuations, i.e. 
\EQ 
\phi(x,t) \,=\,\phi_{cl}(x) + \eta(x,t)
\,\,\, \,\,\, , \,\,\, \,\,\, \eta(x,t) \,=\,\sum_{k} e^{i \omega_k t} 
\,\eta_{k}(x) 
\,\,\,, 
\EN 
and in further expanding the Hamiltonian of the theory in powers of 
$\eta$, by keeping only the quadratic terms. As a result of this procedure,
$\eta_{k}(x)$ satisfies the Schr\"{o}dinger equation 
\EQ
\left[-\frac{d^2}{d x^2} + V''(\phi_{cl}) \right] \, \eta_{k}(x)
\,=\, \omega_k^2 \,\eta_k(x) \,\,\,, \label{schrodinger} 
\EN
together with certain boundary conditions. The semiclassical energy 
levels in each sector are then built in terms
of the energy of the corresponding classical solution and the
eigenvalues $\omega_i$ of the stability equation
(\ref{schrodinger}), i.e. 
\EQ 
E_{\{n_i\}} \,=\,E_{cl} + \hbar \,\sum_{k}\left(n_k
+ \frac{1}{2}\right) \, \omega_k + O(\hbar^2) \,\,\,,
\label{tower} 
\EN 
where $n_k$ are non--negative integers. In particular the ground state energy in each 
sector is obtained by choosing all $n_k = 0$ and it is therefore given by\footnote{From 
now on we will fix $\hbar=1$, since it is well known that the semiclassical expansion
in $\hbar$ is equivalent to the expansion in the interaction
coupling.}
\EQ 
E_{0}
\,=\,E_{cl} + \frac{\hbar}{2} \,\sum_{k} \omega_k + O(\hbar^2)
\,\,\,. \label{e0} 
\EN

In summary, to each static finite energy solution of
(\ref{firstorder}) corresponds a tower of quantum energy
eigenstates (\ref{tower}) representing the $0$--particle (vacua)
and $1$--particle (kink), and their excitations. The construction
of the complete Hilbert space, including the $n$--particle sectors 
(for $n \geq 2$), requires to consider time--dependent multi--kink 
and breather solutions with finite energy. Their semiclassical 
quantization can be performed with an appropriate modification of the 
DHN method \cite{DHN}. 

As we have mentioned in the Introduction, the analytic form of the
semiclassical scaling functions for the 2-D QFT's admitting static
kink solutions can be achieved by DHN method suitably adapted to
finite size geometry. In the following we will discuss in details
the results of Sine--Gordon model on a cylindrical geometry which,
as we shall see, admits the simplest technical analysis. On this
geometry --- described by a space variable compactified on a
circle of circumference $R$ and by a time variable $t$ running on
an infinite interval --- the SG model admits quasi--periodic
boundary conditions (b.c.)
\EQ 
\phi(x + R,t) \,=\,\phi(x,t) +
\frac{2n\pi}{\beta} \,\,\,, \label{quasiperiodicbc}
\EN
where the arbitrary winding number $n\in Z$ originates from the invariance
of the potential (\ref{VSG}) under
$\phi\to\phi+\frac{2n\pi}{\beta}$. In particular, since we are
interested in the one--kink sector, which is defined by $n=1$,
we will impose the b.c.
\EQ
\phi(x + R,t) \,=\,\phi(x,t) +
\frac{2\pi}{\beta} \,\,\,. \label{periodicbc}
\EN

The first step for applying the semiclassical method to this
problem is to find the finite size analog of the kink solution, satisfying
now the b.c.'s (\ref{periodicbc}). However, the success in
constructing the scaling functions depends on whether one is able
to solve the corresponding Schr\"{o}dinger equation (\ref{schrodinger})
and to derive an analytical expression for its frequencies $\omega_k$.
It turns out that the semiclassical finite size effects in SG model
are intrinsically related to the simplest ($N=1$) Lam\`{e} equation,
which admits a complete analytical study.

\subsection{SG in infinite volume}\label{SGinfinite}

The semiclassical quantization of the Sine--Gordon soliton in
infinite volume has been performed in \cite{DHN}. We report here
the basic results in order to show how the semiclassical technique
works in the simplest example and also to introduce the quantities
that should be obtained in the IR limit of the forthcoming finite 
volume results.

The classical (anti)soliton
\begin{equation}
\phi_{cl}(x)=\frac{4}{\beta}
\arctan\left(e^{\pm m (x -x_0) }\right)\;
\label{SGkinkinfvol}
\end{equation}
is solution of eq.\,(\ref{firstorder}) with $A = 0$. It connects the two degenerate
vacua $\phi = 0$ and $\phi = \frac{2\pi}{\beta}$ and its classical mass is given
by $M_{\textrm{cl}} = 8\frac{m}{\beta^{2}}$. Plugging the above expression in
(\ref{schrodinger}), this equation can be cast in the hypergeometric form by
using the variable $z=\frac{1}{2}(1+\tanh mx)$, and its solution is expressed as
\begin{equation}
\eta(x)\,=\,\frac{1}{\beta}\,z^{\frac{1}{2}\sqrt{1-\frac{\omega^{2}}{m^{2}}}}\,
(1-z)^{-\frac{1}{2}\sqrt{1-\frac{\omega^{2}}{m^{2}}}}\,
F\left(2,-1,1+\sqrt{1-\frac{\omega^{2}}{m^{2}}},z\right)\;.
\end{equation}
The corresponding spectrum is given by the discrete value $\omega_{0}^{2} = 0$
(i.e. the zero mode associated to the translation invariance of the theory) 
and by the continuous part $\omega_{q}^{2} = m^{2}(1+q^{2})$, characterized 
by the absence of reflection and by the phase shift $\delta(q) = 2 \arctan(\frac{1}{q})$.

The semiclassical correction to the mass is given by the
difference between the ground state energy of the soliton and the
one of the vacuum, with the addition of a mass counterterm due to
normal ordering of the interaction term in the Hamiltonian:
\begin{equation}
M - M_{\textrm{cl}}\,=\,
\frac{1}{2}\sum_{n}\left[m\sqrt{1+q_{n}^{2}} - \sqrt{k_{n}^{2}+m^{2}}\right]-
\frac{\delta \mu^{2}}{\beta^{2}}
\int\limits_{-\infty}^{\infty}dx\left[1-\cos\beta\phi_{cl}(x)\right]\;,
\label{subinfvol}
\end{equation}
with
\begin{equation}
\delta \mu^{2} \,=\, -\frac{m^{2}\beta^{2}}{8\pi}
\int\limits_{-\infty}^{\infty}
\frac{d k}{\sqrt{k^{2}+m^{2}}}\;.
\end{equation}
The discrete values $q_{n}$ and $k_{n}$ are obtained by defining the
system in a large finite volume of size $R$ with periodic boundary
conditions:
\begin{equation}
2 n\pi \, = \, k_{n} R \,=\, m \,q_{n}\,R + \delta(q_{n})\;.
\end{equation}
Sending $R\rightarrow\infty$ and computing the integrals, one
finally obtains the semiclassical quantum correction to the mass
of the kink
\begin{equation}
M\,=\,\frac{8m}{\beta^{2}}-\frac{m}{\pi}\;.
\label{piccoloh}
\end{equation}
As it is well known, the exact solution of the quantum
Sine--Gordon model shows that the coupling constant $\beta^2$
renormalises as \cite{Jevicki} 
\EQ 
\label{dressing}\beta^2
\rightarrow \gamma \,=\,\frac{\beta^2}{1 - \beta^2/8\pi} \,\,\,.
\EN 
Moreover, the exact quantum mass of the soliton is given by $M
= \frac{8 m}{\gamma}$, which coincides with the above expression
(\ref{piccoloh}). The equality of the semiclassical and the exact
result for the soliton mass is a remarkable property of the SG
model, on which we will come back in the following Sections.

\subsection{Ground state energy regularization in finite volume}\label{secte0reg}

As shown by eq.\,(\ref{e0}), quantum corrections to energy levels
are given by the series on the frequencies $\omega_n$. However,
this series is generally divergent (this is the usual UV
divergence in field theory) and a criterion is needed to
regularize it. It is quite instructive to consider the simplest
example where such divergence occur, i.e. in the calculation of
the ground state energy ${\cal E}_{0}^{\text{vac}}(R)$ of the
vacuum sector of the theory on a cylindrical geometry of
circumference $R$. This can be constructed by implementing the DHN
procedure for one of the constant solutions, for instance
$\phi_{\text{cl}}^{\text{vac}} = 0$, imposing periodic boundary
conditions for the corresponding fluctuations
$\eta^{\text{vac}}(x)$. Obviously, what comes out is nothing
else but the Casimir energy of a free bosonic field $\phi(x,t)$
with mass $m$. In this case the frequency eigenvalues are fixed to
be
\begin{equation}
\omega_{n}\,=\,\sqrt{p_{n}^{2}+m^{2}}\,\,\,,
\end{equation}
with $p_{n} = 2\pi n/R$ and $n=0,\pm 1,\pm 2,\ldots $.

The ground state energy has to be regularized by subtracting its
infinite--volume continuous limit: this ensures in fact the proper
normalization of this quantity, expressed by 
\EQ 
\lim_{R \rightarrow \infty} {\cal E}_0^{\text{vac}}(R) \,=\,0\,\,\,. 
\EN
The ground state energy at a finite volume is therefore defined by
\begin{equation}
{\cal E}_{0}^{\text{vac}}(R) \,=\,
\frac{1}{2}\sum\limits_{n=-\infty}^{\infty}\sqrt{\left(\frac{2 \pi
n}{R} \right)^{2}+m^{2}}\,
-\,\frac{1}{2}\int\limits_{-\infty}^{\infty}dn\,\sqrt{\left(\frac{2
\pi n}{R}\right)^{2}+m^{2}} \,\,\,.
\end{equation}
Isolating the zero mode, it can be conveniently rewritten as
\begin{equation}
{\cal E}_{0}^{\text{vac}}(R) \,=\, \frac{m}{2} +
\frac{2\pi}{R}\sum
\limits_{n=1}^{\infty}\sqrt{n^{2}+\left(\frac{r}{2\pi}
\right)^{2}}\,
-\,\frac{2\pi}{R}\int\limits_{0}^{\infty}dn\,\sqrt{n^{2} +
\left(\frac{r}{2\pi}\right)^{2}} \;,
\end{equation}
where $r \equiv m R$. Since the divergence of the series is due to the large $n$
behaviour of the first two terms in the expansion
\EQ
\sqrt{n^{2}+\left(\frac{r }{2\pi} \right)^{2}}\,\simeq\,
n+\frac{1}{2}\left(\frac{r}{2\pi}\right)^{2}\frac{1}{n} +
{\cal O}\left(\frac{1}{n^2}\right)\;,
\EN
we begin our calculation by subtracting and adding these divergent terms to it:
\begin{eqnarray}
S(r) \,\equiv
\,\sum\limits_{n=1}^{\infty}\sqrt{n^{2}+\left(\frac{r }{2\pi}
\right)^{2}}&=&\sum\limits_{n=1}^{\infty}\left\{\sqrt{n^{2} +
\left(\frac{r}{2\pi}\right)^{2}}-n-\frac{1}{2}\left(\frac{r}{2\pi}\right)^{2}
\frac{1}{n}\right\} + \nonumber \\
&+&\,\sum\limits_{n=1}^{\infty}n\,
+\,\frac{1}{2}\left(\frac{r}{2\pi}\right)^{2}\sum\limits_{n=1}^{\infty}\frac{1}{n}
\,\,\,.
\label{subseriess}
\end{eqnarray}
The first series in the right hand side of the above expression is
now convergent, whereas the last two terms should be coupled to
the analogous ones coming from the integral, whose divergencies have
to be handled in strict correspondence with those coming from the
series. Hence, by subtracting and adding the leading divergence to
the integral
\begin{eqnarray}
I(r) \,& \equiv\,&
\int\limits_{0}^{\infty}dn\,\sqrt{n^{2}+\left(\frac{r }{2\pi}
\right)^{2}} = \nonumber \\
& = & \int\limits_{0}^{\infty}dn\left\{\sqrt{n^{2}+\left(
\frac{r}{2\pi}
\right)^{2}}-n\right\}\,+\,\int\limits_{0}^{\infty}dn\,n\;,
\label{intsub1}
\end{eqnarray}
we can combine the last term in this expression with
the one in (\ref{subseriess}) and implement the well
known regularization
\EQ
\sum_{n=0}^{\infty} n - \int_0^{\infty} n \,dn
\,=\, \lim_{\alpha\rightarrow 0}\left[ \sum_{n=0}^{\infty}
n\,e^{-\alpha n} - \int_0^{\infty} n \,e^{-\alpha n} \,dn \right]
\,=\, -\frac{1}{12} \,\,\,.
\label{firstsubtraction}
\EN
However, the first term in (\ref{intsub1}) still contains a subleading
logarithmic divergence, as it can be seen by explicitly computing the
integral by using a cut-off $\Lambda$, in the limit $\Lambda \rightarrow \infty$
\begin{equation}\label{intsub2}
\int\limits_{0}^{\Lambda}dn\left\{\sqrt{n^{2}+\left(\frac{r }{2\pi}
\right)^{2}}-n\right\}
= \frac{1}{2}\left(\frac{r}{2\pi}\right)^{2}\ln
2\Lambda\,+\,\frac{1}{4}\left(\frac{r}{2\pi}\right)^{2}\,-\,
\frac{1}{2}\left(\frac{r}{2\pi}\right)^{2}\ln\frac{r}{2\pi}
\,\,\,.
\end{equation}
This divergence can be cured by subtracting and adding the term
$\frac{1}{2} \left(\frac{r}{2\pi}\right)^2 \ln \Lambda$. By
combining this last term with its analogous in the series we have
\EQ 
\lim_{\Lambda\to\infty} \left( \sum_{n=1}^{\Lambda}
\frac{1}{n} \,-\,\ln \Lambda \right) \,=\, \gamma_E \,\,\,,
\label{secondsubtraction} 
\EN 
where $\gamma_{E}$ is the Euler-Mascheroni constant, while the 
remaining part of (\ref{intsub2}) with the above subtraction 
is now finite.

Collecting the above results, the finite expression of the ground state energy
on a cylinder is then given by
\begin{equation}
{\cal E}_{0}^{\text{vac}}(R)\,=\,
\frac{1}{R}\left[-\frac{\pi}{6}+\frac{r}{2} +
\frac{r^{2}}{4\pi}\left(\ln\frac{r}{4\pi}+\gamma_{E}-
\frac{1}{2}\right) +\sum_{n=1}^{\infty}\left(\sqrt{(2 \pi n)^{2} +
r^{2}}-2\pi n - \frac{r^{2}}{4\pi n}\right)\right]\;.
\label{Casimiro}
\end{equation}
It is now easy to see that (\ref{Casimiro}) fulfills the
requirement of modular invariance of the theory, which imposes
its equality with the TBA expression \cite{km}
\begin{equation}
{\cal E}_{0}^{\text{vac}}(R) \,=\,-\frac{\pi c(r)}{6 R}\;,
\label{Casimir}
\end{equation}
where
\begin{equation}
c(r) \,=\, -\frac{6 r}{\pi^{2}}\,\int_{0}^{\infty}d\theta\,
\cosh\theta\,\ln\left(1-e^{-r\cosh\theta}\right)\;.
\end{equation}
In fact, this integral formula can be expressed in terms of Bessel
functions, which admit a series representation that directly leads
to (\ref{Casimiro}) (see Ref.\,\cite{km}). For this theory we
obviously have $c(0) = 1$. Moreover, one can also check that the
above regularization scheme ensures the agreement between the $R$
and $L$ channel calculations of the finite expression of the
one--point functions $\langle \phi^{2k} \rangle$ \cite{lm}. The
interested reader can find the simplest example of these
calculations in Appendix \ref{appreg}.

Finally, it is worth to note that the result (\ref{Casimiro}) can
also be obtained by using a simpler prescription which automatically
includes the subtraction of the various divergencies, fastening
the calculation. This consists in ignoring the divergent part of
the integral, keeping only its finite part, and in regularizing
the divergent series as
\begin{eqnarray}
\label{z(-1)}\left.\sum_{n=1}^{\infty} n\;\right|_{\text{reg}}&=&-\frac{1}{12}
\label{riem}\,\,\,, \\
\label{z(1)}\left.\sum_{n=1}^{\infty}\frac{1}{n}\;\right|_{\text{reg}}
& = &\gamma_{E}+\ln\frac{r}{2\pi} \label{magic} \,\,\,.
\end{eqnarray}
Formula (\ref{z(-1)}) is the standard regularization of the
Riemann zeta function $\zeta(-1)$, where $\zeta(s)
\,=\,\sum_{n=1}^{\infty} \frac{1}{n^s}\;$, and usually corresponds
to normal ordering with respect to the infinite volume vacuum
(see, for instance, \cite{birrel}, chapter 4). On the contrary,
the regularization of the second series is a--priori ambiguous due
to its logarithmic divergence, and its finite value (\ref{magic})
was chosen according to the above discussion.

\section{Scaling functions on the cylinder}\label{sectscaling}
\setcounter{equation}{0}

We will now develop a complete semiclassical scheme to analyse the
energy of the quantum state in SG model containing one soliton on
the cylinder. This can be achieved by applying the DHN method to
an appropriate kink background.

\subsection{Properties of the periodic kink solution}\label{periodickinkk}

\begin{figure}[h]
\hspace{3cm} \psfig{figure=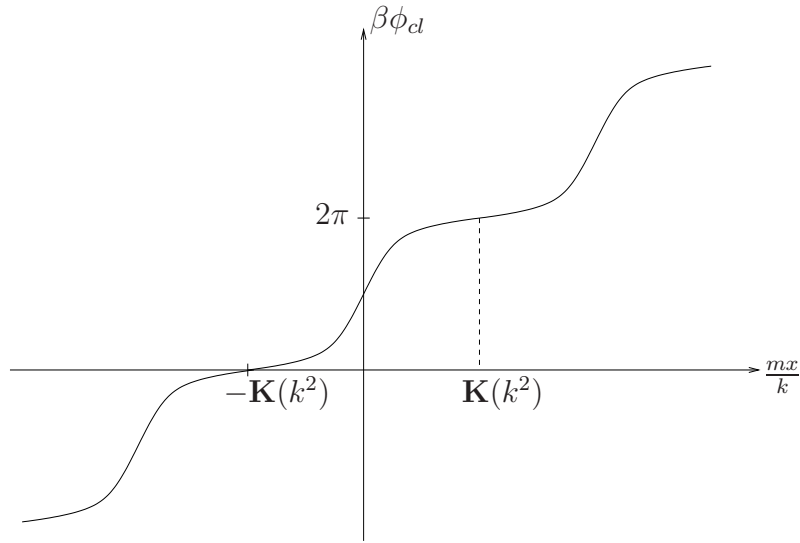,height=7cm,width=10cm}
\vspace{0.1cm} \caption{Solution of eq. (\ref{firstSG}) with $A >
0$ and $x_{0}=0$.} \label{figSGam}
\end{figure}

In order to identify a kink on the cylinder, we have to
look for a static finite energy solution of the SG model
satisfying the quasi--periodic boundary condition
(\ref{periodicbc}). For the first order equation
\begin{equation}
\frac{1}{2}\left(\frac{\partial \phi_{cl}}
{\partial x}\right)^{2} \,=\, \frac{m^{2}}{\beta^{2}}
\left(1-\cos\beta\phi_{cl} + A \right) \,\,\,
\,\,\,
\label{firstSG}
\end{equation}
a solution which has this property can be found for $ A > 0$. It can be
expressed as
\begin{equation}
\phi_{cl}(x)\,=\,\frac{\pi}{\beta} + \frac{2}{\beta}\,
\textrm{am}\left(\frac{m (x - x_0)}{k},k^{2}\right)\;,
\qquad k^{2} \,=\,\frac{2}{2+A}\,\,\,,
\label{SGam}
\end{equation}
provided the circumference $R$ of the cylinder is identified with
\begin{equation}
R \,=\,\frac{2}{m} \,k\,\textbf{K}\left(k^{2}\right)\;\,,
\label{sizeper}
\end{equation}
where $\textbf{K}(k^{2})$ denotes the complete elliptic integral of the first
kind\footnote{The definition and basic properties of $\textbf{K}(k^2)$ and the
Jacobi elliptic amplitude$\,\text{am}(u,k^{2})\,$ can be found in Appendix
\ref{appell}.}. The parameter $x_{0}$ in (\ref{SGam}) represents the kink's
center of mass position, and its arbitrariness is due to the translational
invariance of the theory around the cylinder axis. The behaviour of (\ref{SGam})
as a function of the real variable $x$ is shown in Figure \ref{figSGam}.

\psfrag{phicl(x)}{$\beta\phi_{cl}$} \psfrag{2 pi}{$2\pi$}
\psfrag{K(k^2)}{$\textbf{K}(k^{2})$}
\psfrag{-K(k^2)}{$-\textbf{K}(k^{2})$}
\psfrag{x}{$\hspace{0.2cm}\frac{m x}{k}$}

The function (\ref{SGam}) has been first proposed in \cite{takoka}
and interpreted as a crystal of solitons in the sine-Gordon theory
in infinite volume. In our finite volume case, instead, (\ref{SGam}) 
has to be seen as a single soliton defined on a cylinder of circumference 
$R$ (given by eq.\,(\ref{sizeper})), while its quasi-periodic oscillations 
represent winding around the cylinder. As shown in eq.\,(\ref{sizeper}), 
there is an explicit relation between the size of the system and the 
integration constant $A$. It is easy to see that the infinite volume solution 
(\ref{SGkinkinfvol}) is consistently recovered from (\ref{SGam}) in the limit $A 
\rightarrow 0$, i.e. when $R$ goes to infinity.

The classical energy of the kink on the cylinder is given by
\begin{equation}
{\cal E}_{cl}(R) \,=\, \int\limits_{-R/2}^{R/2}dx
\left[\frac{1}{2} \left(\frac{\partial \phi_{cl} }{\partial
x}\right)^{2} +
\frac{m^{2}}{\beta^{2}}(1-\cos\beta\phi_{cl})\right] \,=\,
\frac{8m}{\beta^{2}}\left[\frac{\textbf{E}(k^{2})}{k} +
\frac{k}{2}\left(1-\frac{1}{k^{2}}\right)\textbf{K}(k^{2})\right]\;,
\label{classenper}
\end{equation}
where $\textbf{E}(k^{2})$ is the complete elliptic integral of the
second kind. In the $R\rightarrow\infty$ limit (which corresponds
to $k'\rightarrow 0$, with $(k')^{2} \equiv 1-k^{2}$), ${\cal
E}_{cl}(R)$ approaches exponentially the correct value $M_{\infty}
= \frac{8m}{\beta^2}$. This can be seen expanding $\textbf{E}$ and
$\textbf{K}$ for small $k'$ (see Appendix B), and expressing the
result in terms of $mR$, which can be itself expanded in $k'$ in
virtue of the relation (\ref{sizeper}):
$$
e^{-m R}\,=\,\frac{1}{16}(k')^{2}+\cdots\;.
$$
Hence the large $R$ expansion of the classical energy is
\begin{equation}
\label{SGclassenexp} {\cal E}_{cl}(R) \,=\, M_{\infty} +
\frac{32}{\beta^{2}}\,m\, e^{-mR} + O\left(e^{-2 m R}\right)\;.
\end{equation}
We will comment more on the interpretation of this result in Section \ref{UVIR}.

Similarly, one can derive the behaviour of ${\cal E}_{cl}(R)$ for
small $r=mR$, which corresponds to the limit $A \rightarrow
\infty$ (or $k^2 \rightarrow 0$): 
\EQ 
{\cal E}_{cl}(R)
\,=\,\frac{2 \pi}{R}\,\frac{\pi}{\beta^{2}} + m\frac{r}{\beta^2} -
m\left(\frac{r}{2 \pi}\right)^3 \, \frac{\pi}{2 \beta^2} + \cdots
\label{smallLcl} 
\EN 
This formula will be relevant in the discussion of the UV properties 
of the scaling functions presented in Sect.\ref{UVIR}.

Before moving to the quantization of the kink--background
(\ref{SGam}), it is worth to mention that another simple kind
of elliptic function, which solves eq.\,(\ref{firstSG}) for
$-2 < A < 0$, was also proposed in \cite{takoka} and interpreted as a
crystal of solitons and antisolitons in the infinite volume SG. 
This background corresponds as well to a kink on the cylinder geometry 
but satisfying the \textit{antiperiodic} boundary conditions 
\[
\phi(x + R,t) \,=\,\frac{2\pi}{\beta}-\phi(x,t) \,\,\,.  
\]
The associated form factors were obtained in \cite{finvolff}. Although the quantization 
of this second kink solution is technically similar to the one of (\ref{SGam}) 
presented in the next section, it displays however some different interpretative 
features that justify its discussion in a separate publication \cite{preparation}.

\subsection{Semiclassical quantization in finite volume}\label{finvolquant}

The application of the DHN method to the periodic kink
(\ref{SGam}) requires the solution of eq.\,(\ref{schrodinger}) for
the quantum fluctuations $\eta_{\omega}$, which in this case takes
the form
\begin{equation}
\left\{\frac{d^{2}}{d
\bar{x}^{2}}+k^{2}\left(\bar{\omega}^{2}+1\right)-2
k^{2}\,\textrm{sn}^{2}(\bar{x},k^2)
\right\}\eta_{\bar{\omega}}(\bar{x}) \,=\, 0\;,
\label{SGsemiclper}
\end{equation}
where $\textrm{sn}(\bar{x},k^2)$ is the Jacobi elliptic function
defined in Appendix \ref{appell}, and we have introduced the
rescaled variables
\begin{equation}
\bar{x}\,=\,\frac{m x}{k}\,,\qquad \bar{\omega}\,=\,\frac{\omega}{m}\;.
\end{equation}
Due to the periodic properties of $\phi_{cl}(x)$ expressed by
eq.\,(\ref{SGam}), the boundary condition (\ref{periodicbc})
translates in the requirement for $\eta_{\bar{\omega}}(\bar{x})$
\EQ 
\eta_{\bar{\omega}}\left(\bar{x} + \frac{m R}{k}\right) \,=\,
\eta_{\bar{\omega}}(\bar{x}) \,\,\,. \label{periodiceta} 
\EN
Eq.\,(\ref{SGsemiclper}) can be cast in the so--called Lam\'e
form, which admits the two linearly independent solutions
\begin{equation}\label{lisol}
\eta_{\pm a}(\bar{x}) \,=\,\frac{\sigma(\bar{x}+i \textbf{K}'
\pm a)}{\sigma(\bar{x} + i \textbf{K}')}\;e^{\mp\,\bar{x} \,\zeta(a)}\;,
\end{equation}
where the auxiliary parameter $a$ is defined as a root of the
equation 
\EQ 
{\cal P}(a) \,= \,
\frac{2-k^{2}}{3}-k^{2}\bar{\omega}^{2} \,\,\,. \label{parametera}
\EN 
The Weierstrass functions ${\cal P}(u)$, $\zeta(u)$ and
$\sigma(u)$ are defined in Appendix \ref{lame}, where the Lam\'e
equation and its relation with (\ref{SGsemiclper}) are discussed
in detail.

\begin{figure}[h]
\hspace{3cm} \psfig{figure=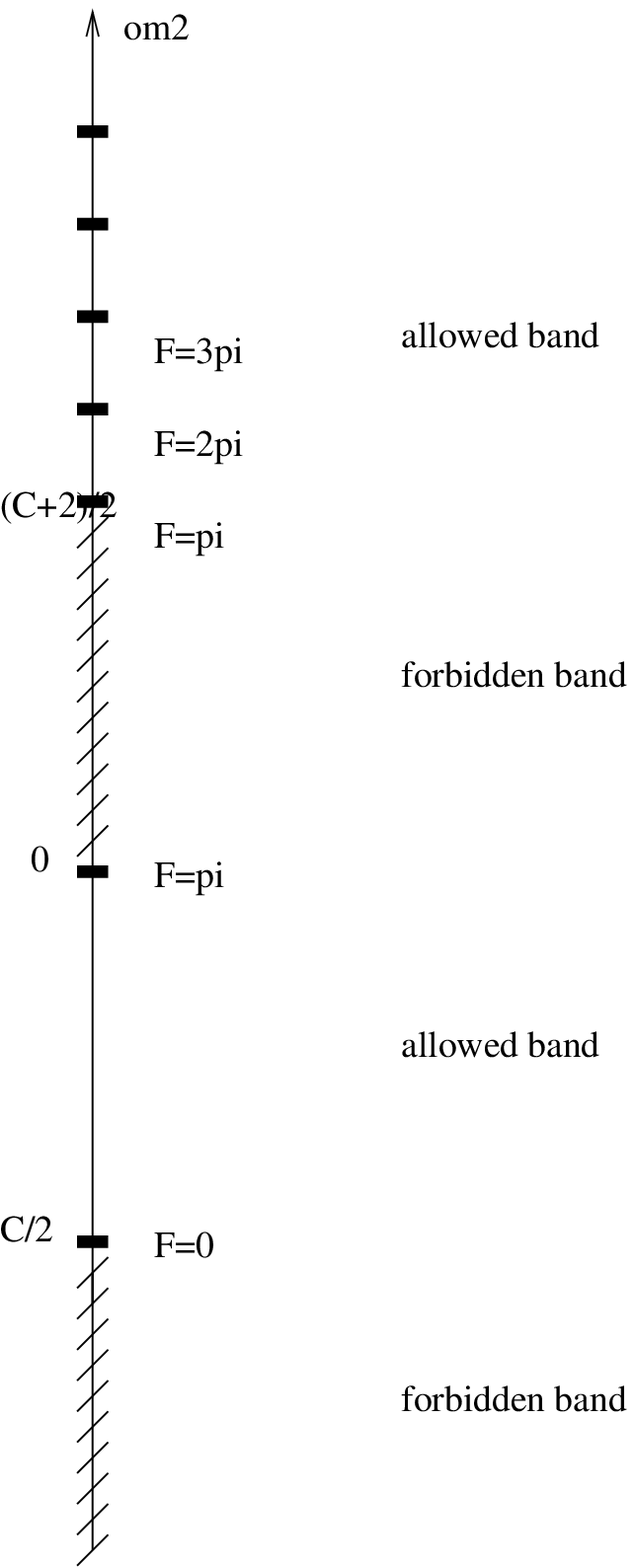,height=7cm,width=10cm}
\vspace{0.1cm} \caption{Spectrum of eq.\,(\ref{SGsemiclper})}
\label{spectrumSGper}
\end{figure}

As it is usually the case for a Schr\"odinger--like equation with
periodic potential, the spectrum of eq.\,(\ref{SGsemiclper}) has a
band structure, determined by the properties of the Floquet
exponent
\begin{equation}\label{floquetper}
F(a) \,=\, 2 i\left[\textbf{K}\, \zeta(a) -
a\,\zeta(\textbf{K})\right] \,\,\,,
\end{equation}
which is defined as the phase acquired by $\eta_{\pm a}$ in
circling once the cylinder
\[
\eta_{\pm a}(\bar{x} + 2 \textbf{K}) \,=\,e^{\pm i F(a)} \,
\eta_{\pm a}(\bar{x}) \,\,\,.
\]
We have two allowed bands for real $F(a)$, i.e. 
\EQ 
0 < \bar{\omega}^{2} < \frac{1}{k^{2}}-1 \,\,\,\,\,\,\,\,
\makebox{and} \,\,\,\,\,\,\,\, \bar{\omega}^{2} > \frac{1}{k^{2}}
\,\,\,, 
\EN 
and two forbidden bands for $F(a)$ complex, i.e. 
\EQ
\bar{\omega}^{2} < 0 \,\,\,\,\,\,\,\, \makebox{and}
\,\,\,\,\,\,\,\, \frac{1}{k^{2}} - 1 < \bar{\omega}^{2} <
\frac{1}{k^{2}} \,\,\,. 
\EN 
The band $0 < \bar{\omega}^2 < \frac{1 - k^2}{k^2}$ is described by 
$a = \textbf{K} + i y$, where $y$ varies between $0$ and $\textbf{K}'$ 
and, correspondingly, $F(a)$ goes from $0$ to $\pi$. The other allowed band 
$\bar{\omega}^2 > \frac{1}{k^2}$ corresponds instead to $a = i y$ and, by varying
$y$, $F(a)$ goes from $\pi$ to infinity, as it is shown in Fig.\,
\ref{spectrumSGper}.

\psfrag{om2}{$\bar{\omega}^{2}$}
\psfrag{(C+2)/2}{$\frac{1}{k^{2}}$}
\psfrag{0}{\hspace{-0.8cm}$\frac{1}{k^{2}}-1$} \psfrag{C/2}{$0$}
\psfrag{F=0}{$F=0$}
\psfrag{F=pi}{$F=\pi$}\psfrag{F=2pi}{$F=2\pi$}\psfrag{F=3pi}{$F=3\pi$}
\psfrag{allowed band}{allowed band}\psfrag{forbidden
band}{forbidden band}

By imposing the periodic boundary conditions (\ref{periodiceta})
on the fluctuation $\eta(\bar{x})$, one selects the values of
$\bar{\omega}^{2}$ for which the Floquet exponent is an even
multiple of $\pi$, thus making the spectrum of eq.\,(\ref{SGsemiclper}) 
discrete. These eigenvalues are $\bar{\omega}_{0}^{2}=0$, which is the 
zero mode associated with translational invariance and has multiplicity one, 
and the infinite series of points
\begin{equation}
\bar{\omega}_{n}^{2} \,\equiv \,\frac{1}{k^{2}}\,
\left[\frac{2-k^{2}}{3} - {\cal P}(i y_{n})\right]\;
\label{omeganper}
\end{equation}
with multiplicity two, placed in the band $\bar{\omega}^{2} > \frac{1}{k^{2}}$,
with $y_{n}$ determined by the equation
\begin{equation}
F(iy_{n}) \,=\, 2\textbf{K}\,i\,\zeta(i y_{n})+2
y_{n}\,\zeta(\textbf{K}) \,=\, 2 n\pi\;, \qquad\quad n=1,2,\ldots
\label{ynper}
\end{equation}

In the IR limit ($A \rightarrow 0 $) the above spectrum goes to the one related
to the standard background (\ref{SGkinkinfvol}): the allowed band $0 < \bar{\omega}^{2}
< \frac{1}{k^{2}}-1$, in fact, shrinks to the eigenvalue $\bar{\omega}_{0}^{2} = 0$,
while the other allowed band $\bar{\omega}^{2} > \frac{1}{k^{2}}$ merges in the
continuous part of the spectrum $\bar{\omega}_{q}^{2} = 1+q^{2}$.

\begin{figure}[h]
\begin{center}
\psfig{figure=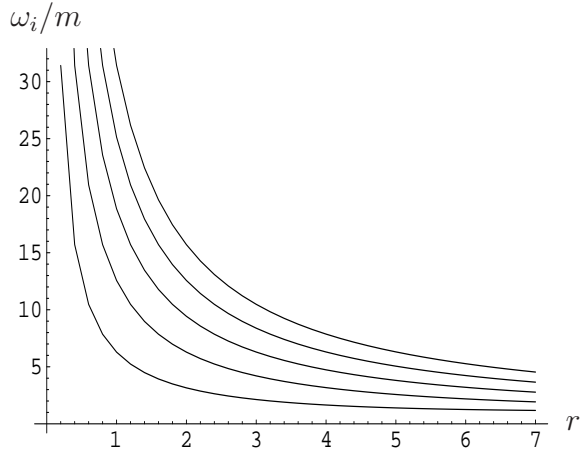,height=6cm,width=8cm} \caption{The first
few levels defined in (\ref{omeganper})}\label{figomegai}
\end{center}
\end{figure}

It is useful to note that, although the $R$ dependence of the frequencies 
(\ref{omeganper}) is quite implicit, since it passes through the inversion of 
eq.\,(\ref{sizeper}), nevertheless these are analytic functions of $R$ and 
it is extremely simple to plot them. The corresponding curves, shown in 
Figure \ref{figomegai}, provide an important piece of information, since 
they are nothing else but the energies of the excited states with 
respect to their ground state ${\cal E}_0(R)$.

\psfrag{omega1}{$\omega_{i}/m$}\psfrag{ell}{$r$}

\begin{figure}[h]
\begin{center}
\psfrag{omega1}{$\omega_{1}/m$}\psfrag{ell}{$r$}
\psfig{figure=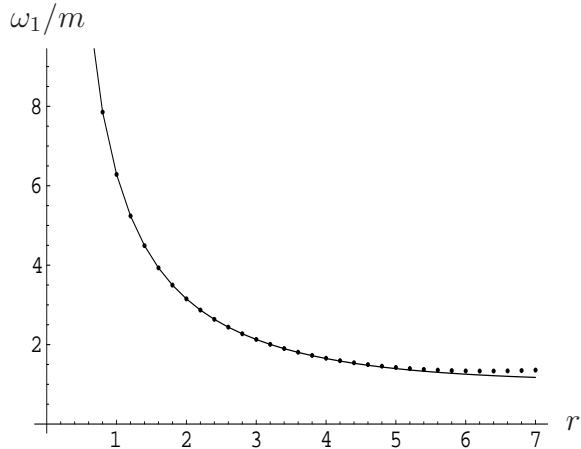,height=6cm,width=8cm}
\end{center}
\caption{Comparison between the exact energy level $\omega_{1}/m$
given by (\ref{omeganper}) (continuous line) and the approximate
expression (\ref{omeganexp}) (dotted line).}\label{figcompar}
\end{figure}

To complete the analysis, it remains then to determine the finite
volume ground state energy ${\cal E}_0(R)$ of the kink sector. 
In analogy with the infinite volume case (see
eq.\,(\ref{subinfvol})), this is defined by 
\EQ 
{\cal E}_0(R) \,=\,
{\cal E}_{cl}(R) +  \sum_{n=1}^{\infty} \omega_n(R) - \frac{\delta
\mu^{2}}{\beta^{2}}\int\limits_{-R/2}^{R/2}dx
\left[1-\cos\beta\phi_{cl}\right] - {\cal E}_0^{\textrm{vac}}(R)
\,\,\,. \label{grstatefinal} 
\EN 
Before commenting in detail each of these terms, let's focus first on the 
main problem in deriving a closed expression for ${\cal E}_0(R)$, which 
consists in the evaluation of the infinite sum on the frequencies $\omega_n(R)$
or, better, in isolating its finite part. We need therefore a
method for solving the transcendental equation (\ref{ynper}) for
$y_n(k^2)$ in order to make the expression (\ref{omeganper}) for
the frequencies $\omega_n(k^2)$ explicit. As we have already seen for 
the classical energy, two kinds of expansion are possible, one in the 
elliptic modulus $k$ and the other in the complementary modulus $k'$, 
which are efficient approximation schemes in the small and large $r$ 
regimes, respectively. Here for simplicity we only present the small $r$ expansion.
By taking into account the series expansion in $k$ for $\textbf{K}$, $\zeta(u)$ and
${\cal P}(u)$ (see Appendices \ref{appell} and \ref{lame}), we are
led to look for a solution of eq.\,(\ref{ynper}) in the form 
\EQ
y_n(k^2) \,=\,\sum_{s=0}^{\infty} (k^2)^s \,y_n^{(s)} \,\,\,. 
\EN
Here we give the result for the first few coefficients
$y_n^{(s)}$, $s=0,1,2$:
\begin{eqnarray}
y_n^{(0)} & = & \textrm{arctanh}\frac{1}{2 n} \,\,\,,\nonumber \\
y_n^{(1)} & = & \frac{1}{4} \; y_n^{(0)} \,\,\,,\\
y_n^{(2)} & = & \frac{9}{64}\; y_n^{(0)} - \frac{n}{16 (4 n^2
-1)^2}\,\,\,, \nonumber
\end{eqnarray}
which are those relevant in the analysis of the UV properties of the
scaling function in Sect.\,\ref{UVIR}. As a consequence, we obtain
the following simple expression for the frequencies:
\EQ
\frac{\omega_n}{m} \,=\, \frac{2 n}{k}\,\left[1 - \frac{k^2}{4} -
\frac{k^{4}}{64}\;\frac{20 n^2 -9}{4 n^2 -1} + O(k^6) \right]
\,\,\,. \label{simpleomegan}
\EN
Comparing it order by order with the small--$k$ expansion of eq.\,(\ref{sizeper})
\EQ
r \,=\, mR \,=\, \pi \, k \,
\left[1+\frac{k^{2}}{4}+\frac{9}{64}\,k^{4}+O(k^{6})\right]\;,
\label{ellexpansion}
\EN
we finally obtain the explicit $R$--dependence
\begin{equation}\label{omeganexp}
\frac{\omega_{n}(R)}{m}\,=\,\frac{2\pi}{r}
\,n+\left(\frac{r}{2\pi}\right)^{3} \,\frac{n}{4n^{2}-1} + \ldots
\end{equation}
It is worth noting that this series expansion in $r$, which can be
easily extended up to desired accuracy, efficiently approximates
the exact energy levels also for rather large values of the
scaling variable. Fig.\,\ref{figcompar} shows a numerical
comparison between the first energy level given by (\ref{omeganper})
and its approximate expression (\ref{omeganexp}), and for the
higher levels it is possible to see that the agreement is even
better.

With the above analysis, the sum over frequencies in the ground
state energy (\ref{grstatefinal}) takes the form
\begin{equation}
\sum\limits_{n=1}^{\infty}\,\frac{\omega_{n}(R)}{m}\,=\,\frac{2\pi}{r}
\,\sum\limits_{n=1}^{\infty}\,n+\left(\frac{r}{2\pi}\right)^{3}
\,\sum\limits_{n=1}^{\infty}\,\frac{n}{4n^{2}-1} + \ldots\;
\end{equation}
As we will see below, the subtraction of counterterm and
vacuum energy in (\ref{grstatefinal}) leads to the cancellation of
all the divergencies, producing a finite expression for the ground 
state energy in the kink sector.

Moving now to the analysis of the remaining terms in
(\ref{grstatefinal}), a similar series expansion can be easily
performed on each of them. The classical energy ${\cal E}_{cl}(R)$,
given in eq.\,(\ref{classenper}), has already be treated
in this way in eq.\,(\ref{smallLcl}). The finite volume counterterm
(C.T.), where the one--loop mass renormalisation is given by
\EQ
\delta
\mu^{2}\,=\,\left.-\frac{m^{2}\beta^{2}}{8\pi}\,
\frac{2\pi}{R}\sum_{n=-\infty}^{\infty}
\frac{1}{\sqrt{m^{2}+\frac{(2 n \pi)^{2}}{R^{2}}}}\right.\,\,\,
\EN
and $\phi_{cl}$ is given by (\ref{SGam}), takes the explicit
form
\EQ 
\text{C.T.}\,=\, m \,
\left[k\textbf{K}\left(k^{2}\right)-\frac{\textbf{K}\left(k^{2}\right)-
\textbf{E}\left(k^{2}\right)}{k}\right]
\left.\sum\limits_{n=-\infty}^{\infty}\frac{1}{\sqrt{(2n\pi)^{2}+r^{2}}}\right.
\;. \label{countfinite}
\EN 
The first terms of its expansion in $R$ are then
\begin{equation}
\frac{\text{C.T.}}{m}\,=\,
\frac{1}{4}\,+\,\frac{r}{4\pi}\left.\sum\limits_{n=1}^{\infty}\frac{1}{n}\right.
\,-\,\frac{r^{2}}{32\pi^{2}}\,-\,\frac{1}{4}\left(\frac{r}{2\pi}\right)^{3}
\left.\sum\limits_{n=1}^{\infty}\left(\frac{1}{n}+\frac{1}{n^{3}}\right)\right.
+ \ldots \;,
\end{equation}
Finally, the vacuum energy ${\cal E}_0^{\textrm{vac}}(R)$ is the
one precisely computed in Sec.\,\ref{secte0reg}. Since its role is
to cancel certain divergencies present in the other terms of
${\cal E}_0(R)$, in complete analogy with the infinite volume case
(see eq.\,(\ref{subinfvol})), we will now consider its \lq\lq
naive" formulation, given by
\begin{equation}
\frac{{\cal E}^{\text{vac}}_{0}(R)}{m}\, =
\,\frac{1}{2m}\sum\limits_{n=-\infty}^{\infty}\sqrt{\left(\frac{2n\pi}{R}\right)^{2}+m^{2}}
\,=\,\frac{1}{2}\,+\,\frac{2\pi}{r}
\left.\sum\limits_{n=1}^{\infty}n\right.\,+\,\frac{r}{4\pi}
\left.\sum\limits_{n=1}^{\infty}\frac{1}{n}\right.
\,-\,\frac{1}{8}\left(\frac{r}{2\pi}\right)^{3}
\left.\sum\limits_{n=1}^{\infty}\frac{1}{n^{3}}\right. + \ldots
\end{equation}
Hence, in the final expression for the ground state energy all the
divergent series present in the sum over frequencies, in the
counterterm and in the vacuum energy cancel out, and one obtains
\begin{equation}\label{grstatefinalexp}
\frac{{\cal E}_{0}(R)}{m} \,=\,
\,\frac{2\pi}{r}\,\frac{\pi}{\beta^{2}}\,-\,\frac{1}{4}\,+
\,\frac{1}{\beta^{2}}\,r
\,-\,\frac{1}{8}\,\left(\frac{r}{2\pi}\right)^{2}\,-\,
\left(\frac{r}{2\pi}\right)^{3} \left[\frac{1}{8}\, \zeta(3)
-\frac{1}{4} (2\,\log 2 -1) - \frac{\pi}{2\beta^{2}}\right] +
\ldots\;,
\end{equation}
where we have used \cite{GRA}
\[
\sum_{n=1}^{\infty} \frac{2 n^2 -1}{8 n^3 (4 n^2 -1)} \,=\,
\frac{1}{8}\, \zeta(3) -\frac{1}{4} (2 \,\log 2 -1) \,\,\,
\]
in order to evaluate explicitly the coefficient of the $r^{3}$ term.

Repeating the above calculations, one can also easily write the
finite expressions of the excited energy levels (\ref{tower}),
whose series expansion in $r$ is given by
\begin{eqnarray}\label{eiexp}
\frac{{\cal E}_{\{k_{n}\}}(R)}{m} & = &
\frac{2\pi}{r}\,\left(\frac{\pi}{\beta^{2}} +
\sum\limits_{n}k_{n}\,n\right)
\,-\,\frac{1}{4}\,+\,\frac{1}{\beta^{2}}\,r
\,-\,\frac{1}{8}\,\left(\frac{r}{2\pi}\right)^{2}\,+\, \\
&& \hspace{-3mm} -\, \left(\frac{r}{2\pi}\right)^{3}
\left[\frac{1}{8} \zeta(3) -\frac{1}{4}(2 \,\log 2 -1)
-\frac{\pi}{2\beta^{2}} +
\sum\limits_{n}k_{n}\,\frac{n}{4n^{2}-1}\right] + \ldots \nonumber
\end{eqnarray}
where $\{k_{n}\}$ is a set of integers defining a particular
excited state of the kink.

\subsection{UV--IR correspondence}\label{UVIR}

The semiclassical quantization of the periodic kink (\ref{SGam})
provides us with analytic expressions, albeit implicit, for the
scaling functions in the kink sector for arbitrary values of the
scale $r = m R$. These quantities control analytically the
interpolation between the Hilbert spaces of the ultraviolet (UV) and
infrared (IR) limiting theories. It is worth to note that, although
we obtain them in the framework of a particle--like description proper
of the IR limit, the kink background (\ref{SGam}) is intrinsically
formulated on a finite size, leading to the possibility of extracting
UV data. Hence, it is important to check whether our scaling functions
reproduce both the expected results for the IR ($r \rightarrow \infty$)
and UV ($r \rightarrow 0$) limits.

\vspace{0.5cm}

Concerning the IR behaviour, we have seen in the previous Sections
that in the $R \rightarrow \infty$ limit all the quantities in
exam, i.e. the classical solution, its classical energy and the
stability frequencies, correctly reach their asymptotic values.
In addition, it is also possible to perform a simple and interesting
analysis of the first correction to the kink mass for large $R$.
According to L\"{u}scher's analysis \cite{luscher}, the mass of a
particle in a large but finite volume has to approach exponentially
its asymptotic value, in a way controlled by the scattering data of
the infinite volume theory. Restricting for simplicity our analysis 
to the leading term in $\beta$ in the kink mass (which, in our approch, 
is simply given by the classical energy), we have then to compare the 
expansion presented in (\ref{SGclassenexp}) with the term that dominates
L\"{u}scher's formula for small $\beta$. This is given by
\cite{luscher,klassen}
\begin{equation}
M(R)-M_{\infty}\,=\,-m_{b}\sin u_{kb}^{k} \,R_{k b k}\;e^{-m_{b}
\sin u_{kb}^{k} R} + \cdots \;, \label{lusch}
\end{equation}
where $R_{k b k}$ is the residue (multiplied by $-i$) of the
kink--breather $S$-matrix on the pole at $\theta=iu_{kb}^{k}$,
i.e.
\EQ
R_{k b k} \,=\,-i \textrm{Res}\, S_{kb}(\theta = i
u_{kb}^k) \,\,\,.
\label{residues}
\EN
Using the kink--breather $S$--matrix \cite{zamzam}
\EQ
S_{k b}(\theta) \,=\,\frac {\sinh \theta + i \cos\frac{\gamma}{16}}
{\sinh \theta - i \cos\frac{\gamma}{16}} \,\,\,\,\,\, ,
\,\,\,\,\,\, \gamma
\,=\,\frac{\beta^2}{1-\beta^2/8\pi}
\EN
and selecting its $s$-channel pole $\theta^* = i u_{kb}^k = i \left(\frac{\pi}{2} +
\frac{\gamma}{16}\right)$, we find
\EQ
R_{k b k} \,=\, - 2\,\textrm{cotg} \frac{\gamma}{16} \,\,\,.
\EN
Substituting in (\ref{lusch}), for small $\beta^2$ we have
\EQ
M(R) - M_{\infty}\,=\,m\,\frac{32}{\beta^2}\,e^{-r} + \cdots \;,
\EN
which therefore reproduces eq.\,(\ref{SGclassenexp}). It is a remarkable
fact that the classical energy alone, being the leading term in
the mass for $\beta^2 \rightarrow 0$, contains the IR scattering
information which controls the large--distance behaviour of ${\cal E}_0(R)$.

\vspace{0.5cm}

The UV behaviour for $r \rightarrow 0$ of the ground state energy
$E_0(R)$ of a given off--critical theory is known to be related
instead to the Conformal Field Theory (CFT) data
$(\Delta,\bar{\Delta},c)$ of the corresponding critical theory and
to the bulk energy term as \cite{cardy}
\EQ
E_0(R)
\,\simeq\,\frac{2\pi}{R} \left(\Delta + \bar{\Delta} -
\frac{c}{12}\right) + B R + \cdots\;
\label{UVCFT}
\EN
where $c$ is the central charge, $\Delta + \bar{\Delta}$ is the lowest
anomalous dimension in a given sector of the theory and $B$ the
bulk coefficient.

For the Sine--Gordon model the bulk energy term is given by
\cite{ddv,bulkterm}
\begin{equation}
B\,=\, 16\,\frac{m^{2}}{\gamma^{2}}\,\tan\frac{\gamma}{16}\; ,
\label{bulktermm}
\end{equation}
while its UV limit is described by the CFT given by the gaussian action
with $c=1$
\EQ
{\cal A}_{\textrm{CFT}} \,=\,\frac{1}{2}\,g \int d^2 x\;
\partial_{\mu} \phi \,\partial^{\mu} \phi \;,
\label{gaussian}
\EN 
where the free bosonic field is compactified on a circle of
radius ${\cal R}$. The various sectors of this CFT are labelled by
two integers, $s$ and $n$: $s$ is the momentum index, while $n$ is
the winding number, related to the boundary condition imposed on
$\phi$ 
\EQ 
\phi(x + R,t) \,=\,\phi(x,t) + 2\pi n\,{\cal R} \;.
\label{winding} 
\EN 
Let $\mid s,n \rangle$ be the states carrying the lowest anomalous 
dimension in each sector. They are created by the vertex operators 
\cite{ginsparg} 
\begin{equation*}
 V_{s,n}(z,\bar{z}) \,=\,:\, \exp\left[i \alpha_{s,n}^+ \varphi(z)
+ i \alpha_{s,n}^- \bar{\varphi}(\bar{z})\right] \,:\;,
\end{equation*}
i.e.
\EQ \mid
s,n\rangle \,=\,V_{s,n}(0,0) \,\mid \textrm{vac}\rangle \,\,\,,
\EN 
where
\begin{eqnarray*}
\alpha_{s,n}^{\pm} & = &\frac{s}{{\cal R}}\pm 2\pi g\,n {\cal R} 
\,\,\,\,; \\
\phi(x,t) & = & \varphi(z) + \bar{\varphi}(\bar{z}) \,\,\,\,\,.
\nonumber
\end{eqnarray*}
Their conformal dimensions are given by
\begin{equation}
\Delta_{s,n}  \,=\, 2\pi g\left(\frac{s}{4\pi g\,{\cal
R}}+\frac{1}{2}\,n {\cal R}\right)^{2} \;,\qquad
\bar{\Delta}_{s,n} \,=\, 2\pi g\left(\frac{s}{4\pi g\,{\cal
R}}-\frac{1}{2}\,n {\cal R}\right)^{2} \,\,\,.
\end{equation}
The vacuum sector is described by $s = n = 0$, with 
$\Delta_{vac} + \bar{\Delta}_{vac} = 0$.
If we now define ${\cal R}=\frac{1}{\sqrt{g}\,\beta}$ and fix the 
normalization constant to the value\footnote{Note that the usual
normalization adopted in the CFT literature is instead
$g = \frac{1}{4\pi}$.} $g=1$, then the kink sector in SG, defined 
by the boundary condition (\ref{periodicbc}), naturally corresponds to
the sector characterized by $s=0,n=1$, in which the lowest anomalous 
dimension is
\EQ 
\Delta_{0,1} + \bar{\Delta}_{0,1} \,=\,
\frac{\pi}{\beta^2} \,\,\,.
\EN
The conformal vertex operator $V_{0,1}$ has been put in exact 
correspondence with the soliton--creating operator of SG in
Ref.\,\cite{BerLec}.

The question to be addressed now is whether the small $r$ expansion of
${\cal E}_0^{vac}(R)$ and ${\cal E}_0(R)$ given by
eqs.\,(\ref{Casimiro}) and (\ref{grstatefinalexp}) reproduces, in
semiclassical approximation, the above data controlling the UV
limit of SG model.

For the vacuum sector, comparing (\ref{Casimiro}) with
(\ref{UVCFT}), we correctly obtain $c = 1$ and $\Delta_{vac} =
\bar{\Delta}_{vac} =0$. We do not expect, however, to obtain the
bulk term $B$ relative to SG model by looking at (\ref{Casimiro}),
simply because the semiclassical expression of the ground state
energy in the vacuum sector applies equally well to any theory
which has a quadratic expansion near the vacuum state. Namely,
apart from the value of the mass $m$, eq.\,(\ref{Casimiro}) is a
universal expression that does not refer then to SG model.

The kink scaling function (\ref{grstatefinalexp}) has instead a richer
structure. The obtained scaling dimension
\begin{equation}
\Delta+\bar{\Delta} \,=\,\frac{\pi}{\beta^{2}}\;
\end{equation}
is the expected one for the soliton-creating operator in Sine-Gordon while
the central charge contribution $c=1$ is absent, simply because in
(\ref{grstatefinalexp}) we have subtracted the vacuum ground state energy from
the kink one\footnote{The value $c=1$, coming out from the
regularization of the leading term of the series on the
frequencies (\ref{omeganexp}), is in fact exactly cancelled by the same term
in the vacuum energy.}. Moreover, the bulk coefficient
$B = \frac{m^{2}}{\beta^{2}}$ present in (\ref{grstatefinalexp})
correctly reproduces the semiclassical limit of the exact one, given in
eq.\,(\ref{bulktermm}). In principle, this bulk term should be present in
all the energy levels, included the ground state energy in the vacuum sector, 
but its non--perturbative nature makes impossible to see it in the
semiclassical expansion around the vacuum solution, which is in fact purely
perturbative. Hence it is not surprising that to extract the bulk energy term
we have to look at the kink ground state energy, in virtue of the non--perturbative
nature of the corresponding classical solution.

Finally, the expression (\ref{eiexp}) for the excited energy
levels explicitly show their correspondence with the conformal
descendants of the kink ground state. In fact, their anomalous
dimension is given by
\begin{equation}
\Delta_{\{k_{n}\}}+\bar{\Delta}_{\{k_{n}\}}=\frac{\pi}{\beta^{2}}+
\sum\limits_{n}k_{n}\,n\;. 
\end{equation}

\vspace{0.5cm}

The successful check with known UV and IR asymptotic
behaviours confirms the ability of the semiclassical results to
describe analytically the scaling functions of SG model in the
one--kink sector. It would be interesting to further test them at
arbitrary values of $r$ through a numerical comparison with the
results of \cite{ddv, RavaniniSG} in an appropriate range of
parameters. This was not pursued here because the results presently 
available in the literature were obtained for values of $\beta$ which 
are beyond the semiclassical regime and moreover the energy levels 
were plotted as functions of a different scaling variable, i.e. the 
one defined in terms of the kink mass. We hope however to come back 
to this problem in the future.

\section{Form factors and correlation functions}\label{sectff}
\setcounter{equation}{0}

The semiclassical scaling functions, derived in Sect.\,\ref{sectscaling},
provide an important information about the finite size effects in SG model.
As in the infinite volume case, however, the complete description of the
finite volume QFT requires to find, in addition to the energy eigenvalues
(\ref{eiexp}), the kink form factors and the correlation functions
of local operators. This section is devoted to the analysis of this problem,
i.e. to the determination of the finite volume form factors and the
corresponding spectral functions.

\subsection{Infinite volume form factors}

It is useful to initially recall some basic definitions and
results concerning semiclassical form factors for the SG model 
in infinite volume. As mentioned in Sect.\,\ref{sectsm}, the 
relation between kink solutions and form factors was established 
by Goldstone and Jackiw \cite{goldstone}, who showed that the 
matrix element of the field $\phi$ between two
asymptotic one--kink states is given, at leading order in the
semiclassical regime, by the Fourier transform of the classical
solution describing the kink itself (see also \cite{pathint} for
further developments). This remarkable result, however, had the
drawback of being formulated non--covariantly in terms of the kink
space--momenta. It was refined in \cite{finvolff} with a covariant
formulation in terms of the rapidity variable $\theta$ of the
kink, defined in terms of its energy and momentum as $E = M\cosh
\theta\,,\; p = M \sinh \theta$. In the semiclassical regime,
there are moreover further simplifications: in fact, the mass of
the kink can be approximated by its classical energy $M \simeq
M_{cl} = \frac{8m}{\beta^{2}}$ whereas its rapidity can be assumed
to be very small, i.e. $\theta = O(\beta^{2})$, thus obtaining
$\;E\simeq M_{cl}\,,\; p\simeq M_{cl}\theta\;$. Hence, the refined
form of Goldstone and Jackiw result is given by 
\EQ 
\langle \theta_1 \mid \phi(0) \mid \theta_2\rangle \,=\, M_{cl} \,
\int_{-\infty}^{\infty} da \,e^{-i M_{cl} (\theta_{1}-\theta_{2})
a} \, \phi_{cl}(a) \,\,\,, \label{ffkink} 
\EN 
where $|\,\theta_{i}\rangle$ are asymptotic one--kink states. Moreover,
it is also possible to prove that the form factor of an operator
expressible as a function of $\phi$ is given by the Fourier
transform of the same function of $\phi_{cl}$. For instance, the
form factor of the energy density operator $\varepsilon$ can be
computed performing the Fourier transform of $\varepsilon_{cl}(x)
= \frac{1}{2}\left(\frac{d\phi_{cl}}{dx}\right)^{2} +
V[\phi_{cl}]$. With this covariant formulation, the matrix element
(\ref{ffkink}) can be continued to the crossed channel, and from
its pole structure one can easily extract the spectrum of the
bound states of the theory, even in non--integrable cases, as
discussed in \cite{dsgmrs,finvolff}.

\begin{figure}[h]
\begin{tabular}{p{8cm}p{8cm}}
\psfrag{Fs}{$F\hspace{3cm} \beta=0.1$}\psfrag{t}{$\theta$}
\psfig{figure=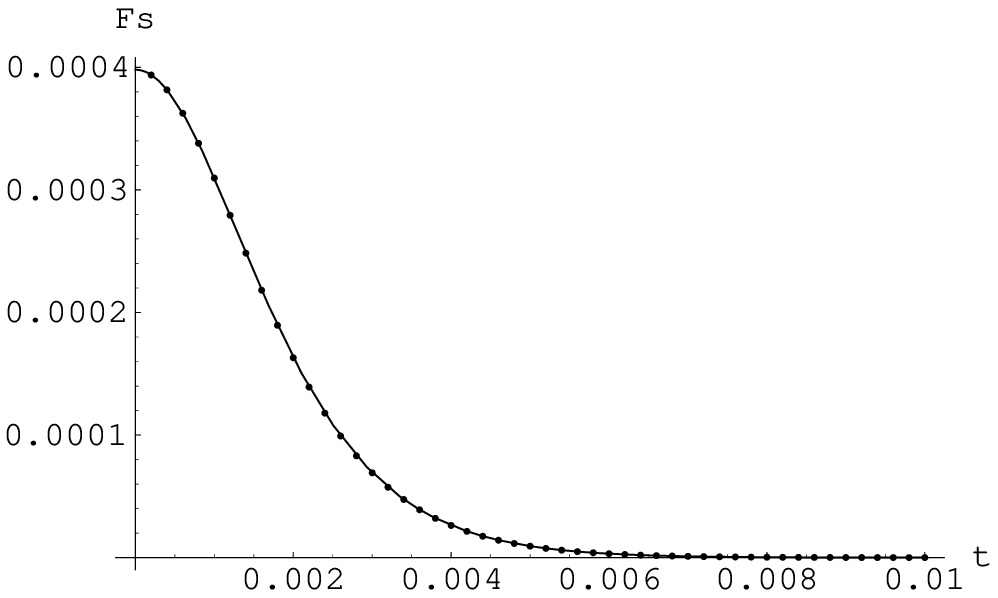,height=5cm,width=7cm}&
\psfrag{Fs}{$F\hspace{3cm} \beta=0.5$}\psfrag{t}{$\theta$}
\psfig{figure=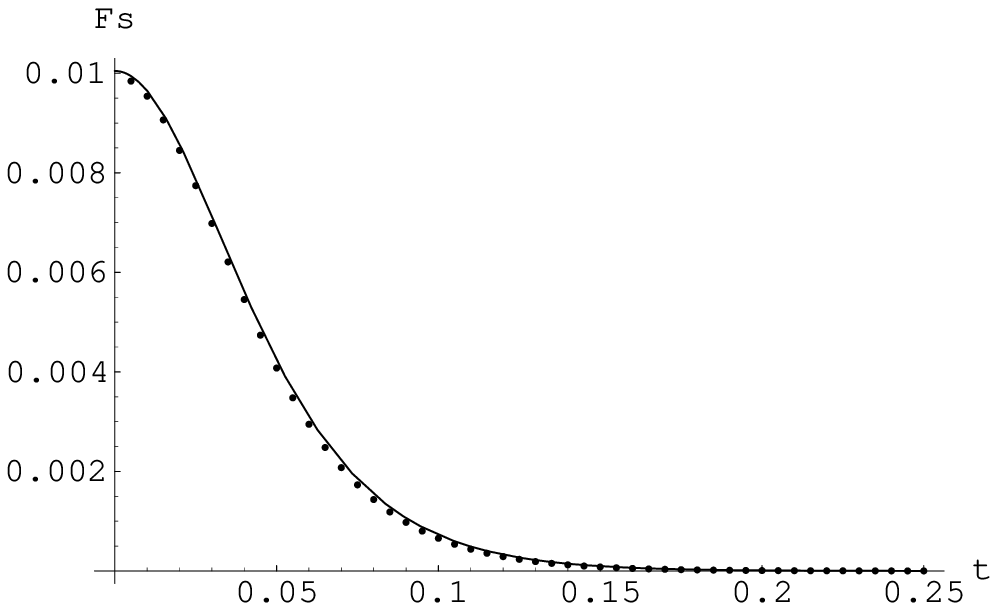,height=5cm,width=7cm}
\end{tabular}
\caption{Comparison between the exact function $F$ given by
(\ref{Fex}) (continuous line) and its semiclassical approximation
(\ref{Fsem}) (dotted line), at $\beta=0.1$ and
$\beta=0.5$.}\label{figcheckquant}
\end{figure}

\begin{figure}[h]
\begin{tabular}{p{8cm}p{8cm}}
\psfrag{Fs}{$F\hspace{3cm} \beta=1\qquad(a)$}\psfrag{t}{$\theta$}
\psfig{figure=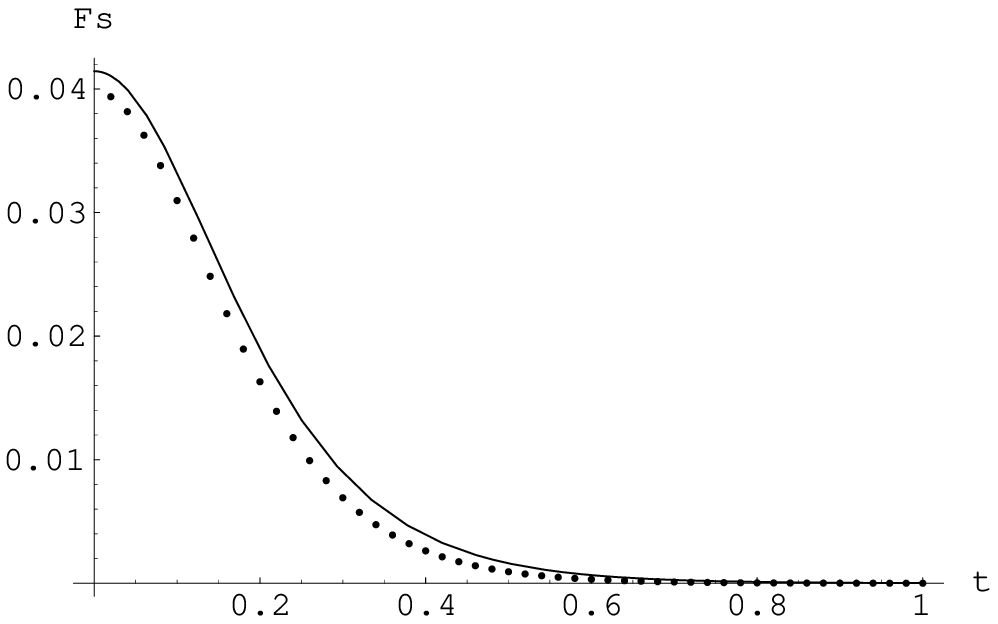,height=5cm,width=7cm}&
\psfrag{Fs}{$F\hspace{3cm} \beta=1\qquad(b)$}\psfrag{t}{$\theta$}
\psfig{figure=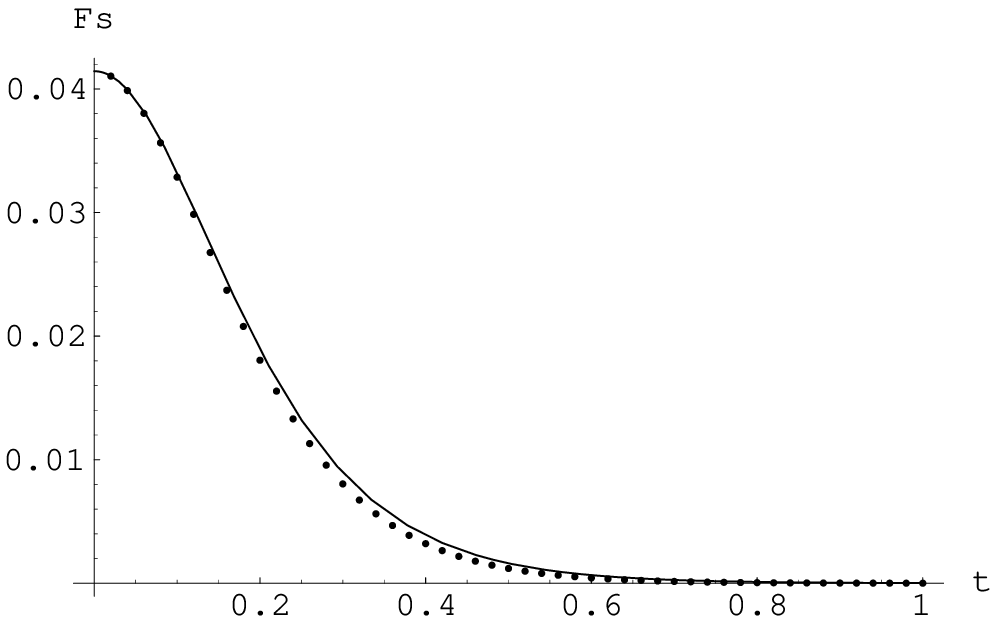,height=5cm,width=7cm}
\end{tabular}
\caption{Comparison, at $\beta=1$, between (a) the exact function
$F$ given by (\ref{Fex}) (continuous line) and its semiclassical
approximation (\ref{Fsem}) (dotted line), (b) the exact function
$F$ given by (\ref{Fex}) (continuous line) and its semiclassical
approximation (\ref{Fsem}) with the substitution
$\beta^{2}\to\gamma$ (dotted line).}\label{figcheckquantdress}
\end{figure}

In \cite{finvolff} we have checked that the form factor
(\ref{ffkink}) obtained from the infinite volume kink
(\ref{SGkinkinfvol}) reproduces the semiclassical limit of the
exact one, derived in \cite{karowski}. Here we would like to present a more quantitative comparison
which permits to conclude that formula (\ref{ffkink}), though
proven under the semiclassical assumption of small coupling and 
small rapidities, remarkably extends its validity to finite values of 
the coupling and to a large range of the rapidities.  
Consider, for instance, the form factor of the energy operator
(up to a normalization $N$) $F(\theta)\,=\,N\langle\,\theta_{2}\,
|\,\epsilon(0)\,|\,\theta_{1}\,\rangle$, whose semiclassical and
exact expressions are given, respectively, by
\begin{eqnarray}\label{Fsem}
F_{\text{semicl.}}(\theta)&=&\hspace{5mm}\frac{\theta}{2}
\hspace{3mm}\frac{1}{\sinh \left[\frac{4\pi}{\beta^{2}} \,\theta\right]}
\\\label{Fex}
F_{\text{exact}}(\theta)&=&\sinh\frac{\theta}{2}\hspace{3mm}
\frac{1}{\sinh\frac{\theta}
{2\xi}} \hspace{3mm} G(\theta)
\end{eqnarray}
where $\xi\,=\,\gamma/8\pi$ and
\EQ
\label{G}
G(\theta)\,=\,\text{exp}\left[\int\limits_{0}^{\infty}\frac{dt}{t}\;
\frac{\sinh\frac{t}{2}(1-\xi)}{\sinh\frac{t}{2}\,\xi\;\;\cosh\frac{t}{2}}\;
\frac{\sin^{2}\frac{\theta \,t }{2\pi}}{\sinh t}\right] \,\,\,.
\EN
Fig.\,\ref{figcheckquant} shows how, for small values of the coupling, 
the agreement between the two functions is very precise for the whole range 
of the rapidity. Furthermore, the discrepancy between exact and semiclassical
formulas at larger values of $\beta$ can be simply cured, in our example, 
by substituting the bare coupling $\beta^2$ with its renormalized
expression $\gamma$ into the semiclassical result (\ref{Fsem}), as 
shown in Fig.\,\ref{figcheckquantdress}. Hence we can conclude that the 
monodromy factor (\ref{G}), which is the relevant quantity missing in our 
approximation, does not play a significant role in the quantitative evaluation 
of the form factor even for certain finite values of the coupling\footnote{
It is easy to understand the reason of this conclusion in the above example: 
at small values of $\theta$ we have $G(\theta) \simeq 1$, whereas for $\theta 
\rightarrow \infty$, when $G(\theta)$ may contribute, the whole form factor 
goes anyway to zero. Similar conclusion can be reached for all other form factors 
which vanish at $\theta \rightarrow \infty$.}.  

As we have already mentioned, the exactness (or very high
accuracy, as in this case) of the semiclassical results is a
peculiar feature of SG model in infinite volume, obtained with the
\lq\lq dressing" $\beta^2 \rightarrow \gamma$. An interesting
problem to be studied is whether similar phenomena take place for
the semiclassical scaling functions and form factors in finite
volume as well. An indication on this issue could be found
by extending to finite volume the analysis of higher loop quantum
corrections which, in the semiclassical approach, are obtained 
by keeping cubic (and higher) powers of $\eta$ in the expansion of
$V(\phi_{\text{cl}}+\eta)$ \cite{pathint}.

\subsection{Semiclassical spectral functions on the cylinder}

The generalization of the above construction to the case of finite
volume has been proposed in \cite{finvolff}, where we have shown how
to estimate the leading semiclassical behaviour of the spectral function
on the cylinder under the same hypotheses of the infinite volume case.
In fact, the matrix element
\begin{equation}
f(\theta_{n})\,=\,\langle
p_{n_{2}}|\,\phi(0)\,|p_{n_{1}}\rangle\;,
\end{equation}
of the basic field $\phi$ between two kink eigenstates of the
finite volume hamiltonian can be expressed, at leading order, as
the Fourier transform of the corresponding classical solution:
\begin{equation}\label{ff}
f(\theta_{n}) \,\equiv \,M(R)\,\int\limits_{-R/2}^{R/2}
da\,e^{i\,M(R)\theta_{n} a}\,\phi_{cl}(a)\;,
\end{equation}
\begin{equation}\label{inverseff}
\phi_{cl}(a)\,\equiv\,\frac{1}{R\,M(R)}\,
\sum\limits_{n=-\infty}^{\infty}\,e^{-i\,M(R)\theta_{n} a}
\,f(\theta_{n})\;.
\end{equation}
Here we have denoted the states by the so-called "quasi-momentum"
variable $p_{n}$, which corresponds to the eigenvalues of the
translation operator on the cylinder (even multiples of $\pi/R$),
and we have defined $\theta_{n}$ as the "quasi-rapidity" of the
kink states
\begin{equation}
\frac{2n\pi}{R}\,=\,p_{n}\,=\, M(R)\,\sinh \theta_{n}\, \simeq \,
M(R)\, \theta_{n}\;,
\end{equation}
where $M(R)$ is the classical mass of the finite--volume kink. As
shown in \cite{finvolff}, the crossed channel form factor
can be obtained at leading order via the change of variable
$\theta\rightarrow i\pi-\theta$:
\begin{equation}
F_{2}(\theta_{n})\,=\,\langle\,
0\,|\,\phi(0)\,|\bar{p}_{n_{2}}\,p_{n_{1}}\rangle\,=\,f(i\pi-\theta_{n})\;,
\end{equation}
and the leading terms in the spectral function on the cylinder
are given by
\begin{equation}
\hat\rho(E_{k},p_{k}) \,= \,
2\pi\delta(E_{k})\delta(p_{k})|<0|\,\phi(0)|\,0>|\,^{2} +
\frac{\pi}{4}\, \frac{\delta\left(\frac{E_{k}}{M}-2\right)}
{M^{2}}\,\left|F_{2}\left(
i\pi-\frac{p_{k}}{M}\right)\right|^{2}\,\,\,.
\end{equation}


The procedure described above has been introduced in
\cite{finvolff} for the construction of form factors for the kink
backgrounds in the SG model and the broken $\phi^{4}$ field
theory, both defined on a cylindrical geometry with
\textit{antiperiodic} boundary conditions. In what follows we will
apply it instead to the case of SG model with \textit{periodic} boundary
conditions. The corresponding finite volume form factor (\ref{ff})
can be written in terms of the soliton background (\ref{SGam}):
\begin{equation}\label{SGff}
f(\theta_{n})=M\int \limits_{-R/2}^{R/2}da\,e^{i\,M\theta_{n}
a}\left[\frac{\pi}{\beta}+\frac{2}{\beta}\,\textrm{am}\left(\frac{m
x}{k},k^{2}\right)\right] =
\end{equation}
\begin{equation*}
=\frac{2\pi}{\beta}\left\{\frac{M}{2}\,R\,
\delta_{M\theta_{n},0}-i\,\frac{1-\delta_{M\theta_{n},0}}{\theta_{n}}
\left[\cos\left(M\theta_{n}\,R/2\right)-
\frac{\sin\left(M\theta_{n}\,R/2\right)}{M\theta_{n}R/2}\right]
+i\,\frac{1}{\,\theta_{n}\,\cosh\left(k\,\textbf{K}'\frac{M}{m}\,
\theta_{n}\right)}\right\}\;.
\end{equation*}
In order to obtain this result one has to compare the inverse
Fourier transform (\ref{inverseff}) with the expansion \cite{GRA}
\begin{equation}
\textrm{am}(u) \,=\,\frac{\pi \,u}{2\textbf{K}}+
\sum\limits_{n=1}^{\infty} \frac{1}
{\,n\,\cosh\left[n\pi\frac{\textbf{K}'}{\textbf{K}}\right]}\,
\sin\left[n\pi\frac{u}{\textbf{K}}\right]\;.
\end{equation}

The form factor (\ref{SGff}) has the correct IR limit\footnote{The
functions $\frac{\cos(x R/2)}{x}$ and $\frac{\sin(x R/2)}{x^{2} R/2}$
can be shown to tend to zero in the distributional sense for
$R\rightarrow\infty$.}, and leads to the following expressions for
$F_{2}(\theta)$ and for the spectral function\footnote{Here we are
considering the matrix elements on the antisymmetric combinations
of kink and antikink.}:
\begin{eqnarray}\nonumber
F_{2}(\theta_{n})&=&\frac{4\pi i}{\beta\, \hat{\theta}_{n}}\,
\left\{\frac{1}{\cosh\left[k\,\textbf{K}'\,\frac{M}{m}\,\hat{\theta}_{n}\right]}\,
\, + \right.\\
&&\left.-\,\left(1-\delta_{\hat{\theta}_{n},0}\right)
\left[\cos\left(M \,\hat{\theta}_{n}\,R/2\right)
- \frac{\sin\left(M \hat{\theta}_{n}\,R/2\right)}
{M \,\hat{\theta}_{n} \,R/2}\right]\right\}\label{SGfinvolf2} \\
\hat\rho(E_{n},p_{n}) & = &
4\pi^{3}\,\delta\left(\frac{E_{n}}{M}-2\right)\frac{1}{\beta^{2}(p_{n})^{2}}
\left\{\frac{1}{\cosh\left[\frac{k\,\textbf{K}'}{m}\,p_{n}\right]} \,\, + \right.
\nonumber \\
&& \left. -
\left(1-\delta_{p_{n},0}\right)
\left[\cos\left(p_{n}\,R/2\right)-\frac{\sin\left(p_{n}\,R/2\right)}
{p_{n}R/2}\right]
 \right\}^{2}\;, 
\label{SGfinvolrho}
\end{eqnarray}
where $\hat\theta = i \pi - \theta$. Note that the finite volume dependence 
of both the form factor (\ref{SGfinvolf2}) and the spectral function (\ref{SGfinvolrho})
is not restricted to the second term only. The $k \,\textbf{K}'(k^{2})\,M(R)$ factor in 
the first term carries the main $R$-dependence, although it is not manifest but 
implicitly defined by eq.\,(\ref{sizeper}).

\vspace{0.5cm}

Another quantity of interest is the two--point function $\langle\, 0\mid
\varepsilon(x)\varepsilon(0)\mid 0\,\rangle$ of the energy density
operator. One can calculate it by evaluating its spectral function
\begin{equation}
\rho_{\varepsilon}(p^{2})=\int\limits_{-R/2}^{R/2}dx\,\langle\,
0\mid \varepsilon(x)\varepsilon(0)\mid 0\,\rangle \, e^{-ip\cdot
x}
\end{equation}
in terms of the form factors of $\varepsilon(x)$, similarly to
what we have done for the SG field $\phi$ above. In order to find
the semiclassical form factor
\begin{equation}
f_{\varepsilon}(\theta_{n})\,=\,\langle
p_{n_{2}}|\,\varepsilon(0)\,|p_{n_{1}}\rangle\;,
\end{equation}
we need to compute the Fourier transform of
\begin{equation}
\varepsilon(\phi_{\text{cl}})\,=\,\frac{2
m^{2}}{\beta^{2}k^{2}}(1+k^{2})-\frac{4
m^{2}}{\beta^{2}}\,\text{sn}^{2}\left(\frac{m x}{k}\right)\;.
\end{equation}
This can be easily obtained from the following expansion
\begin{equation}\label{expffeps}
\text{sn}^{2}u\,=\,\frac{1}{k^{2}\textbf{K}}\left\{\textbf{K}-\textbf{E}-\frac{\pi^{2}}{\textbf{K}}
\sum\limits_{n=1}^{\infty}\frac{n\,\cos\frac{n\pi
u}{\textbf{K}}}{\sinh\frac{n\pi\textbf{K}'}{\textbf{K}}}\right\}\;,
\end{equation}
and we finally have
\begin{equation}
f_{\varepsilon}(\theta_{n})\,=\,M^{2}\left\{\delta_{M\theta_{n},0}\,
+\,\frac{4\pi}{\beta^{2}}\,\frac{\theta_{n}}{\sinh\left(k\,\textbf{K}'\frac{M}{m}\,\theta_{n}\right)}\right\}\;.
\end{equation}
The corresponding semiclassical spectral function is thus given by
\begin{equation}
\hat\rho_{\varepsilon}(E_{n},p_{n}) =
\frac{4\pi^{3}}{\beta^{4}}\,\delta\left(\frac{E_{n}}{M}-2\right)\frac{p_{n}^{2}}
{\sinh^{2}\left(\frac{k\,\textbf{K}'}{m}\,p_{n}\right)}\;.
\end{equation}

It is worth to mention that it is also possible to obtain the two--point functions 
of certain vertex operators $V_{b}^{\pm}(x,t)=e^{\pm i\beta b \phi(x,t)}$ (for
$b = \frac{1}{2},1,\frac{3}{2},2,...$), since the required Fourier
expansion formulas of the type (\ref{expffeps}) are known in these cases \cite{whit}.

\section{Further directions}\label{conclusions}
\setcounter{equation}{0}

In this paper we have shown how the semiclassical
methods can provide an analytic description of finite size
effects in two--dimensional quantum field theories displaying
degenerate vacua. In particular, we have applied these techniques to
study the SG model, quantizing its kink solution on the cylinder. 
The scaling functions of the ground (and excited) states, as well as 
the form factors and two--point functions of different operators, 
allowed us to build the one--kink sector (i.e. $Q_{\text{top}}=\pm 1$) 
of the corresponding Hilbert space of states.

The next step in this program is the extension of the DHN method
to describe the multi--kink states ($Q_{\text{top}}=\pm 2,\pm
3,...$) as well as the non--vacua (\lq\lq breather"--like) part of
the $Q_{\text{top}}=0$ sector. These states are related to certain 
time--dependent solutions of SG model on the cylinder, i.e. to the 
finite volume analog of soliton--soliton, soliton--antisoliton and 
breather solutions. Although more complicated from the technical point of
view, the determination of these classical solutions and the study
of their scaling functions and form factors is a well stated open
problem in the semiclassical framework, which deserves further
attention.

One of the advantages of the semiclassical method is that it works
equally well for both integrable and non--integrable models, if
they admit kink--type solutions. In fact, we have chosen to test
the efficiency of the semiclassical quantization on the example of
SG model, mainly because it leads to the simplest $N=1$ Lam\'e
equation. Static elliptic solutions for other models can be
easily obtained by integrating equation (2.2) with $A\neq 0$ and
appropriate boundary conditions. This was done, for instance in 
\cite{finvolff}, where we have derived the form factors between 
kink states in the broken $\phi^{4}$ model on the cylinder with 
twisted boundary conditions. In this case, the quantization of the 
finite volume kink involves a Lam\'e equation with $N=2$. 
Lam\'e equations with $N>2$ are also expected to enter the quantization 
of other theories. 

Finally, the semiclassical method seems to be suited also for the
description of finite geometries with boundaries, say with Dirichlet or 
Neumann boundary conditions. Due to the physical significance of this 
kind of systems, an interesting problem is the semiclassical computation 
of the relative energy levels, a subject that will be discussed in a forthcoming 
publication  \cite{preparation}. 

\vspace{1cm}

\begin{flushleft}\large
\textbf{Acknowledgements}
\end{flushleft}
This work was partially supported by the Italian COFIN contract
``Teoria dei Campi, Meccanica Statistica e Sistemi Elettronici''
and by the European Commission TMR programme HPRN-CT-2002-00325
(EUCLID). G.S. thanks FAPESP for the financial support. G.M 
would like to thank the Instituto de Fisica Teorica in San Paulo 
for the warm hospitality.

\newpage

\begin{appendix}

\section{Free theory quantization on a finite geometry}\label{appreg}
\setcounter{equation}{0}

Let us consider a free bosonic field $\phi(x,t)$ of mass $m$
defined on a cylinder of circumference $R$, i.e. satisfying the
periodic boundary condition
\EQ
\phi(x + R, t) \,=\,\phi(x,t)
\,\,\,.
\EN
Imposing the equation of motion and the commutation
relation
\EQ
[\,\phi(x,t),\Pi(y,t)\,] \,=\,i \delta_P(x-y) \,\,\,,
\EN
where $\Pi(x,t) = \frac{\partial\phi}{\partial t}(x,t)$ is the
conjugate momentum of the field whereas
\[
\delta_P(x) \,=\,\frac{1}{R} \sum_{n=-\infty}^{\infty} e^{\frac{2
\pi i n}{R}\, x} \,\,\,\,\,\,\,\, , \,\,\,\,\,\,\,\, \delta_P(x+R)
\,=\,\delta_P(x)
\]
is the periodic version of the Dirac delta function, we obtain the
mode expansion of the field $\phi(x,t)$. This is given by
\EQ
\phi(x,t)
\,=\, \sum_{n=-\infty}^{\infty} \frac{1}{2 \omega_n R} 
\left[\,A_n \,e^{i(p_n x - \omega_n t)} + A^{\dagger} \, 
e^{-i (p_n x - \omega_n t)} \,\right] \,\,\,,
\EN
where
\[
[A_n,A^{\dagger}_m] \,=\,\delta_{n,m} \,\,\,,
\]
and 
\EQ 
\omega_n \,=\,\sqrt{p_n^2 + m^2} \,\,\,\,\,\,\,\, ,
\,\,\,\,\,\,\,\, p_n \,=\,\frac{2 \pi n}{R} \,\,\,\, 
n =0,\pm 1,\ldots 
\EN 
Using the above expansion together with the commutation relation 
of $A$ and $A^{\dagger}$, it is easy to compute the propagator of 
the field, given by 
\EQ
\Delta_F(x-x',t-t') \,=\, \langle \phi(x,t) \phi(x',t') \rangle
\,=\, \sum_{n=-\infty}^{\infty} \frac{1}{2\omega_n R}
\,e^{-i[\omega_n (t-t') -p_n (x-x')]} \,\,\,. 
\EN 
The vacuum expectation value of the operator $\phi^2(x,t)$ is then formally
given by 
\EQ 
\langle \phi^2(x,t) \rangle \,=\,\Delta_F(0) \,\,\,
\EN 
and, by translation invariance, is independent from $x$ and
$t$. However this expression is divergent and needs therefore to
be regularized. Analogously to what has been done in the text for
the ground state energy ${\cal E}_0^{\text{vac}}(R)$, we need to 
subtract the corresponding expression in the infinite volume, so 
that the finite quantity, simply denoted by $\phi^2_0(R)$, satisfies 
the usual normalization condition
\[
\lim_{R \rightarrow \infty} \phi^2_0 (R) \,=\,0\,\,\,.
\]
Hence we define 
\EQ 
\phi^2_0 (R) \,=\,  \frac{1}{2 R} \,\sum\limits_{n=-\infty}^{\infty}
\frac{1}{\sqrt{\left(\frac{2 \pi n}{R} \right)^{2}+m^{2}}}\, -\,\frac{1}{2 R}
\,\int\limits_{-\infty}^{\infty} dn \frac{1}{\sqrt{\left(
\frac{2 \pi n}{R} \right)^{2}+m^{2}}} \;. 
\EN 
Isolating its zero mode, the series needs just one subtraction, i.e. 
\EQ 
{\cal S}(r) \,\equiv\,
\sum\limits_{n=1}^{\infty}\frac{1}{\sqrt{n^{2}+\left(\frac{r}{2\pi}
\right)^{2}}} \,=\,
\sum\limits_{n=1}^{\infty}\left\{\frac{1}{\sqrt{n^{2}+\left(\frac{r}{2\pi}
\right)^{2}}}\,-\,\frac{1}{n}
\right\}\,+\,\sum\limits_{n=1}^{\infty}\frac{1}{n}\,\,\,. 
\EN 
($r = m R$). In the above expression, the first series is now
convergent whereas the second series, which is divergent, has to
be combined with a divergence coming from the integral. Indeed we
have 
\EQ 
{\cal I}(r) \,\equiv\,
\int\limits_{0}^{\infty}dn\frac{1}{\sqrt{n^{2}+\left(\frac{r}{2\pi}
\right)^{2}}} \,=\, \lim_{\Lambda\to\infty}\left\{\ln
2\Lambda-\ln\frac{r}{2\pi}\right\}
-\lim_{\Lambda\to\infty}\ln\Lambda+\lim_{\Lambda\to\infty}\ln\Lambda\;,
\EN 
and the last term can be used to compose (\ref{secondsubtraction}). 
Collecting the above expressions, it is now easy to see that 
$\phi^2 _0(R)$ coincides  with the one obtained doing the calculation 
in the other channel, i.e. at a finite temperature. In fact, using 
the results of Ref.\,\cite{lm}, this quantity can be expressed as 
\EQ 
\phi^2 _0(R) \,=\,
\int_{-\infty}^{\infty} \frac{d\theta}{2\pi} \frac{1}{e^{r
\cosh\theta} -1} \,\,\,, 
\EN 
whose expansion in $r$ is given by
\EQ 
\phi^2 _0(R) \,=\, \frac{1}{2 r} + \frac{1}{2\pi}
\left(\log\frac{r}{2\pi} + \gamma_E -\log 2\right) +
\sum_{n=1}^{\infty} \left( \frac{1}{\sqrt{(2 n \pi)^2 + r^2}} -
\frac{1}{2 n \pi}\right) \,\,\,. 
\EN

Also this result could have been directly obtained computing only
the finite part of the integral and using the prescription
(\ref{z(1)}).

\vspace{3mm}

\section{Elliptic integrals and Jacobi's elliptic
functions}\label{appell}
\setcounter{equation}{0}

In this appendix we collect the definitions and basic properties
of the elliptic integrals and functions used in the text.
Exhaustive details can be found in \cite{GRA}.

The complete elliptic integrals of the first and second kind,
respectively, are defined as
\begin{equation}\label{ellint}
\textbf{K}(k^{2})\,=\,\int\limits_{0}^{\pi/2}\frac{d\alpha}
{\sqrt{1-k^{2}\sin^{2}\alpha}}\;,\qquad
\textbf{E}(k^{2})\,=\,\int\limits_{0}^{\pi/2}d\alpha
\sqrt{1-k^{2}\sin^{2}\alpha}\;.
\end{equation}
The parameter $k$, called elliptic modulus, has to be bounded by
$k^{2} < 1$. It turns out that the elliptic integrals are nothing
but specific hypergeometric functions, which can be easily
expanded for small $k$:
\begin{eqnarray}
\textbf{K}(k^{2})&=&\frac{\pi}{2}\;F\left(\frac{1}{2},\frac{1}{2},1;k^{2}\right)=
\frac{\pi}{2}\,\left\{1+\frac{1}{4}\,k^{2}+\frac{9}{64}\,k^{4} + \ldots +
\left[\frac{(2n-1)!!}{2^{n}n!}\right]^{2}k^{2n} + \ldots \right\}\;,\\\nonumber
\textbf{E}(k^{2})&=&\frac{\pi}{2}\;F\left(-\frac{1}{2},\frac{1}{2},1;k^{2}\right)=
\frac{\pi}{2}\,\left\{1-\frac{1}{4}\,k^{2}-\frac{3}{64}\,k^{4}+ \ldots -
\left[\frac{(2n-1)!!}{2^{n}n!}\right]^{2}\frac{k^{2n}}{2n-1} + \ldots \right\}\;.
\end{eqnarray}
Furthermore, for $k^{2}\to 1$, they admit the following expansion
in the so--called complementary modulus $k' = \sqrt{1-k^{2}}$:
\begin{eqnarray}
\textbf{K}(k^{2})&=&
\log\frac{4}{k'}+\left(\log\frac{4}{k'}-1\right)\frac{k'^{2}}{4} 
+ \ldots\;,\\\nonumber
\textbf{E}(k^{2})&=&1+\left(\log\frac{4}{k'} - 
\frac{1}{2}\right)\frac{k'^{2}}{2} + \ldots \;.
\end{eqnarray}
Note that the complementary elliptic integral of the first kind is
defined as 
\EQ
\textbf{K}'(k^{2}) \,=\, \textbf{K}(k'^{2})\;.
\EN

The function $\text{am}(u,k^{2})$, depending on the parameter $k$,
and called Jacobi's elliptic amplitude, is defined through the
first order differential equation
\begin{equation}\label{ellam}
\left(\frac{d\,\text{am}(u)}{du}\right)^{2}\,=\,
1 - k^{2} \sin^{2} \left[\text{am}(u)\right]\;,
\end{equation}
and it is doubly quasi--periodic in the variable $u$:
$$
\text{am}\left(u+2n\textbf{K}+2im\textbf{K}'\right) \,= \, n\pi+\text{am}(u)\;.
$$
The Jacobi's elliptic function $\text{sn}(u,k^{2})$, defined
through the equation
\begin{equation}\label{ellsn}
\left(\frac{d\,\text{sn}u}{du}\right)^{2}
\,=\, \left(1-\text{sn}^{2}u\right)\left(1-k^{2}\text{sn}^{2}u\right)\;,
\end{equation}
is related to the amplitude by $\text{sn}\,u=\sin\left(\text{am}\,u\right)$, 
and it is doubly periodic:
$$
\text{sn}\left(u+4 n\textbf{K}+2im\textbf{K}'\right)\,=\,\text{sn}(u)\;.
$$

\section{Lam\'e equation}\label{lame}
\setcounter{equation}{0}

The second order differential equation
\begin{equation}\label{lameeq}
\left\{\frac{d^{2}}{d u^{2}}-E-N(N+1){\cal P}(u)\right\}f(u) \,=\, 0\;,
\end{equation}
where $E$ is a real quantity, $N$ is a positive integer and 
${\cal P}(u)$ denotes the Weierstrass function, is known under 
the name of $N$-th Lam\'e equation. The function ${\cal P}(u)$ is 
a doubly periodic solution of the first order equation (see \cite{GRA})
\begin{equation}\label{defP}
\left(\frac{d{\cal P}}{du}\right)^{2}\,=\,4\left({\cal P} - 
e_{1}\right)\,\left({\cal P} - e_{2}\right)\,\left({\cal P} - e_{3}\right)\;,
\end{equation}
whose characteristic roots $e_{1},e_{2},e_{3}$ uniquely determine
the half--periods $\omega$ and $\omega'$, defined by
$$
{\cal P}\left(u+2n\omega+2 m\omega'\right)\,=\,{\cal P}(u)\;.
$$

The stability equation (\ref{SGsemiclper}) can be identified with
eq. (\ref{lameeq}) for $N=1$, $u=\bar{x}+i\textbf{K}'$ and
$E=\frac{2-k^{2}}{3}-k^{2}\bar{\omega}^{2}$ in virtue of the
relation between ${\cal P}(u)$ and the Jacobi elliptic function
$\text{sn}(u,k)$ (see formulas 8.151 and 8.169 of \cite{GRA}):
\begin{equation}\label{Psn}
k^{2}\text{sn}^{2}(\bar{x},k) \,=\,{\cal P}(\bar{x}+i\textbf{K}') + 
\frac{k^{2}+1}{3}\;.
\end{equation}
Relation (\ref{Psn}) holds if the characteristic roots of ${\cal P}(u)$ 
are expressed in terms of $k^{2}$ as
\begin{equation}\label{roots}
e_{1}\,=\,\frac{2-k^{2}}{3}\;,
\qquad e_{2}\,=\,\frac{2k^{2}-1}{3}\;,\qquad
e_{3} \,=\,-\frac{1+k^{2}}{3}\;,
\end{equation}
and, as a consequence, the real and imaginary half periods of
${\cal P}(u)$ are given by the elliptic integrals of the first
kind
\begin{equation}\label{halfper}
\omega \,=\,\textbf{K}(k)\;,\qquad \omega' \,=\, i \textbf{K}'(k) \;.
\end{equation}
All the properties of Weierstrass functions that we will use in
the following are specified to the case when this identification
holds.

In the case $N=1$ the two linearly independent solutions of
(\ref{lameeq}) are given by (see \cite{whit})
\begin{equation}
f_{\pm a}(u)\,=\,\frac{\sigma(u\pm a)}{\sigma(u)}\;e^{\mp\,u
\,\zeta(a)}\;,
\end{equation}
where $a$ is an auxiliary parameter defined through ${\cal P}(a)=E$, 
and $\sigma(u)$ and $\zeta(u)$ are other kinds of
Weierstrass functions:
\begin{equation}
\frac{d\,\zeta(u)}{du} \,=\, - {\cal P}(u)\;,\qquad
\frac{d\,\log\sigma(u)}{du} \,=\, \zeta(u)\;,
\end{equation}
with the properties
\begin{eqnarray}\nonumber
&&\zeta(u+2\textbf{K}) \,=\,\zeta(u) + 2 \zeta(\textbf{K})\;,\\\label{zsprop}
&&\sigma(u+2\textbf{K})\,=\,-\,e^{2(u+\textbf{K}) \zeta(\textbf{K})} \sigma(u)\;.
\end{eqnarray}
As a consequence of eq. (\ref{zsprop}) one obtains the Floquet
exponent of $f_{\pm a}(u)$, defined as
\begin{equation}
f(u+2 \textbf{K}) \,=\, f(u)e^{i F(a)}\;,
\end{equation}
in the form
\begin{equation}
F(\pm a) \,=\, \pm 2
i\left[\textbf{K}\,\zeta(a) - a\,\zeta(\textbf{K})\right]\;.
\end{equation}
The spectrum in the variable $E$ of eq. (\ref{lameeq}) with $N=1$
is divided in allowed/forbidden bands depending on whether $F(a)$
is real or complex for the corresponding values of $a$. We have
that $E < e_{3}$ and $e_{2} < E < e_{1}$ correspond to allowed bands,
while $e_{3} < E < e_{2}$ and $E > e_{1}$ are forbidden bands. Note that
if we exploit the periodicity of ${\cal P}(a)$ and redefine
$a\rightarrow a' = a + 2 n \omega+2 m \omega'$, this only shifts $F$
to $F' = F+2 m \pi$.

The function $\zeta(u)$ admits a series representation
\cite{hancock} that will be very useful for our purposes in Sect.
\ref{finvolquant}:
\begin{equation}
\zeta(u)\,=\,\frac{\pi}{2\textbf{K}}\,\cot\left(\frac{\pi u}
{2\textbf{K}}\right)+\left(\frac{\textbf{E}}{\textbf{K}}
+\frac{k^{2}-2}{3}\right)\,u + \frac{2\pi}{\textbf{K}}
\sum\limits_{n=1}^{\infty}
\frac{h^{2n}}{1-h^{2n}}\,\sin \left(\frac{n \pi u}{\textbf{K}}\right)\;,
\end{equation}
where $h = e^{-\pi\textbf{K}'/\textbf{K}}$. The small-$k$ expansion
of this expression gives
\begin{equation}
\label{expzeta}
\hspace{-3.9cm}\zeta(u)\,=\,\left(\cot u + \frac{u}{3}\right)
\,+\,\frac{k^{2}}{12}\left(u - 3 \cot u + 3 u \cot^{2}u\right)\,+
\end{equation}
\begin{equation*}
+\,\frac{k^{4}}{64}\left(-3 u + ( 4 u^{2} - 5 ) \cot u + u \cot^{2}u 
+ 4 u^{2} \cot^{3}u + \sin 2u \right) + \ldots
\end{equation*}
(note that $h\approx \left(\frac{k}{4}\right)^{2}+O\left(k^{4}\right)$). 
A similar expression takes place for ${\cal P}(u)$, by noting that 
${\cal P}(u) = - \frac{d\,\zeta(u)}{du}$.

\end{appendix}

\end{document}